\documentclass[fleqn,usenatbib,useAMS]{mnras}

\usepackage{multicol}
\usepackage{graphicx}
\usepackage{amsmath}
\usepackage{amssymb}	
\usepackage{bm}
\usepackage{multicol}
\usepackage{physics}

\newcommand{\bfv}{\mathbf{v}}

\newcommand{\OmK}{\Omega_\text{K}}

\title[Global 3D simulations of Outer PPDs]{Global Three-Dimensional Simulations of Outer Protoplanetary Disks with Ambipolar Diffusion}

\author[]{Can Cui$^{1}$\thanks{E-mail: \href{mailto:cc795@cam.ac.uk}{cc795@cam.ac.uk}} and 
Xue-Ning Bai$^{2,3}$\thanks{E-mail: \href{mailto:xbai@tsinghua.edu.cn}{xbai@tsinghua.edu.cn}}
\\
$^{1}$DAMTP, University of Cambridge, CMS, Wilberforce Road, Cambridge CB3 0WA, UK \\
$^{2}$Institute for Advanced Study, Tsinghua University, Beĳing 100084, China  \\
$^{3}$Department of Astronomy, Tsinghua University, Beĳing 100084, China}

\pubyear{2021}

\begin{document}
\label{firstpage}
\pagerange{\pageref{firstpage}--\pageref{lastpage}}
\maketitle

\begin{abstract}
The structure and evolution of protoplanetary disks (PPDs) are largely governed by disk angular momentum transport, mediated by magnetic fields. In the most observable outer disk, ambipolar diffusion is the primary non-ideal magnetohydrodynamic(MHD) effect.%PPD gas dynamics is primarily controlled by ambipolar diffusion as the dominant non-ideal magnetohydrodynamic (MHD) effect.
In this work, we study the gas dynamics in outer PPDs by conducting a series of global 3D non-ideal MHD simulations with ambipolar diffusion and net poloidal magnetic flux, using the Athena++ MHD code, with resolution comparable to local simulations. Our simulations demonstrate the co-existence of magnetized disk winds and turbulence driven by the magneto-rotational instability (MRI). While MHD winds dominate disk angular momentum transport, the MRI turbulence also contributes significantly. We observe that magnetic flux spontaneously concentrate into axisymmetric flux sheets, leading to radial variations in turbulence levels, stresses, and accretion rates. Annular substructures arise as a natural consequence of magnetic flux concentration. The flux concentration phenomena show diverse properties with different levels of disk magnetization and ambipolar diffusion. The disk generally loses magnetic flux over time, though flux sheets could prevent the leak of magnetic flux in some cases. Our results demonstrate the ubiquity of disk annular substructures in weakly MRI turbulent outer PPDs, and imply a stochastic nature of disk evolution.
\end{abstract}


\section{Introduction}\label{sec:intro}

Protoplanetary disks (PPDs) are the birth place of planets, where dust and ice particles coagulate into planetesimals, seeding further growth into terrestrial planets and cores of gas giants. PPDs accrete onto the central protostar at typical rate of $\sim10^{-8\pm 1}M_\odot$ yr$^{-1}$ (e.g., \citealt{hartmann_etal98,hh08}), with typical lifetime of several million years \citep{haisch_etal01,mamajek09}. The accretion and dispersal of PPDs are governed by disk gas dynamics, which is also the key to understanding many physical processes in planet formation. In particular, the mechanisms of angular momentum transport shape the structure and govern the global evolution of the disk, and also determine the level and nature of disk turbulence, thereby affecting virtually all stages of planet formation \citep{armitage10}.

In the recent years, the Atacama Large Millimiter/submillimeter Array (ALMA) has been revolutionary that uncovers a diverse range of disk substructures \citep{alma_etal15,andrews20},
particularly annular substructures (e.g \citealp{isella_etal16,andrews_etal16,vanboekel_etal17,avenhaus_etal18,long_etal18,huang_etal18}), spirals (e.g \citealp{benisty_etal15,perez_etal16,huang_etal18b}), and crescent-shaped azimuthal asymmetries (e.g \citealp{vdmarel_etal13,vdmarel_etal20}). These features have triggered
extensive research on their origins. They could be interpreted as signposts of unseen planets embedded in the disk,
reflecting planet formation processes caught in action, while they may simply reflect local concentration of dust that could potentially become precursors to planet formation.

Below, we give a brief overview on the above two aspects of recent development before introducing the objectives of this study.

\subsection{Angular Momentum Transport}

Magnetic fields are known to play a leading role in driving angular momentum transport in PPDs (e.g., see recent reviews by \citealp{turner_etal14,wbf21}). The driving mechanisms can be broadly divided into two categories -- radial transport of angular momentum mediated by the turbulence through the magnetorotational instability (MRI; \citealp{c61,bh91}) and vertical transport of angular momentum through magnetized disk winds \citep{bp82}.

Conventionally, the MRI turbulence is considered as the primary mechanism to transport angular momentum, which requires the perfect coupling between disk gas and magnetic fields. However, the extremely weakly ionized gas in PPDs undermines this coupling \citep{gammie96,wardle07,bai11a}, which is measured by three non-ideal magnetohydrodynamic (MHD) effects -- Ohmic resistivity, Hall effect, and ambipolar diffusion (AD). Among them, Ohmic resistivity and AD are dissipative and effectively stabilize the MRI such that in inner disks the MRI is suppressed by Ohmic resistivity at the midplane (e.g., \citealp{fs03,ilgnernelson08,oishi_etal09}) and by AD at the disk surface \citep{bs13,bai13,gressel_etal15}. In outer disks, AD is the dominant non-ideal MHD effect. The MRI is found to be damped (but not suppressed) by AD at the midplane, but is expected to become more vigorous at the disk surface due to substantially boosted ionization levels by stellar far-ultraviolet (FUV) radiation \citep{pc11,simon_etal13a,simon_etal13b,bai15}.

With the MRI turbulence being largely suppressed or damped, several purely hydrodynamic instabilities have been identified and systematically explored (e.g. \citealp{kb03,nelson_etal13,marcus_etal13,kh14,ly15,cl21}). The onset and outcome of these instabilities sensitively depend on disk thermodynamics, despite that the resulting turbulence level remains relatively low, insufficient to account for the bulk accretion rate in PPDs (see \citealp{lu19} for a recent review).

In the meantime, it is revealed that magnetized disk winds play a major role in driving disk accretion. This has been robustly demonstrated in the inner disk regions in both local \citep{bs13,bai13,bai14,lesur_etal14} and global simulations \citep{gressel_etal15,bai17,wang_etal19,gressel_etal20}, where all three non-ideal MHD effects are relevant and dominate in different vertical heights. As a prerequisite, disk winds requires the disk be threaded by net vertical magnetic flux, which is likely inherited from the star formation process (see, e.g., \citealp{zhao+20} for a review). The efficiency of angular momentum transport is then directly linked to the amount of magnetic flux threading the disk.

In outer disks, however, the situation remains ambiguous. Although high resolution local simulations can appropriately resolve the MRI \citep{simon_etal13a,simon_etal13b,bai15}, the role played by disk winds is uncertain for two reasons. First, local framework ignores disk curvature, hence cannot distinguish between inward and outward radial directions, which is particularly problematic when the disk is turbulent \citep{bs13a}. Second, the disk surface is artificially truncated by the vertical boundaries that is not able to accommodate the wind propagation (\citealp{fromang_etal13}, but see \citealp{riols16}). On the other hand, existing global simulations are either in 2D thus cannot capture the MRI (e.g., \citealp{bs17,bethune_etal17,suriano_etal18,rl20}), or in 3D with resolution too coarse to properly resolve the MRI (e.g., \citealp{bethune_etal17,suriano_etal19,hu_etal19}). Although these studies show efficient wind-driven angular momentum transport, it is unclear that to what extent their results will be undermined due to the lack of third dimension and/or resolution.

Therefore, a fundamental question left unanswered is whether magnetized disk wind can still be launched in outer PPDs, and the relative importance angular momentum by the MRI and the wind.

\subsection{Annular Substructures}

Annular structures are one of the most common type of disk substructures (see review by \citealp{andrews20}). They are most
commonly considered as a consequence of planet-disk interactions (e.g. \citealp{gt79,lp86,lp93}), with significant recent development
to interpret disk observations (e.g. \citealp{bryden_etal99,nelson_etal00,zhu_etal11,kn12,dong_etal17}). This interpretation is recently
corroborated by additional signatures in gas kinematics, suggesting the presence of Jupiter-mass planet responsible of carving deep
gaps in a few systems \citep{teague_etal18,pinte_etal19}. 

While the planet scenario is especially compelling, it does not necessarily apply to all annular substructures observed. In particular, rings and gaps are also discovered during
the embedded Class 0/I phase \citep{se18,nakatani_etal20,segura-cox_etal20,sheehan_etal20} at early stages of disk formation, requiring either rapid formation of massive planets, or alternative, planet-free interpretations.

Indeed, there is extensive recent literature on alternative origins of disk rings.
To name but a few, one set of proposed mechanisms is related to the condensation fronts of various volatile species, which mark a
transition in dust properties (e.g., \citealp{zhang_etal15,okuzuni_etal16,hu_etal19,owen20}). Another set of scenarios is associated
with gas-dust coupling, including secular gravitational instability \citep{sc11,youdin11,ts14,tominaga_etal18,tominaga_etal19},
dust-driven viscous ring-instability \citep{wunsch05,dp18}, and self-induced dust traps \citep{drazkowka_etal16,gonzalez_etal17}.

Forming annular substructures generally requires a redistribution of disk mass, and hence disk angular momentum. Since magnetism
dominates the angular momentum transport in PPDs, any non-smoothness in angular momentum transport associated with magnetic
fields must directly lead to the formation of annular substructures. This leads to another broad category of mechanisms to generate
rings and gaps. These include MRI-induced zonal flows (e.g., \citealp{johansen_etal09,dittrich_etal13,sa14}), gas density
accumulation at the dead-zone boundaries \citep{vt06,flock_etal15}, and spontaneous magnetic flux concentrations \citep{bs14,bethune_etal17,suriano_etal18,suriano_etal19,rl19,rl20}\footnote{ We note that the Hall effect could also lead to spontaneous magnetic flux concentration, but has only been observed  in unstratified simualtions (e.g. \citealp{kunz+13,bethune+16,krapp+18}).
}.

We note that zonal flows and magnetic flux concentration are intimately related \citep{bs14}, which are the main focus of this work.
It was found in local and global disk simulations that through the development of the MRI and launching of magnetized disk winds,
the global distribution of poloidal magnetic flux threading the disk tends to become non-uniform and spontaneously concentrate into
quasi-axisymmetric flux sheet. Regions with stronger magnetic flux transports angular momentum more efficiently and become
depleted (i.e., gaps), whereas mass tends to accumulate in regions in between flux sheets (i.e., rings). 
However, the aforementioned works were conducted either as local simulations, which again suffer from numerical artifacts, or
global simulations that significantly under-resolve the MRI. It is yet unclear whether magnetic flux concentration is a robust
phenomenon, how magnetic flux concentrates at first place, hence the characteristics of the resulting features.

% deplete; evacuate

\subsection{This Work}

In this paper, we conduct global 3D non-ideal MHD simulations of the outer PPDs. Our simulations include net poloidal magnetic flux, with domain size opening up to the polar region to fully accommodate wind launching, while in the meantime feature a numerical resolution comparable to those achieved in local shearing box simulations. As an initial study, we avoid complications from evaluating the diffusion coefficients self-consistently from ionization chemistry and thermodynamics. Instead, our simulations are designed to be scale-free to systematically scan the parameter space by prescribing the strength of AD and level of magnetization. In doing so, we aim to clarify the dominant mechanism governing angular momentum transport in outer PPDs, and in the meantime, we will verify how disks naturally form annular substructures from magnetic flux concentration.

The paper is organized as follows. In \S\ref{sec:method}, we describe numerical methods and simulation setup. In \S\ref{sec:diag}, we list diagnostic quantities to facilitate the analysis of simulation results. The results of fiducial models including gas dynamics and magnetic flux concentration phenomenon are detailed in \S\ref{sec:fi}. We conduct parameter study on the strength of ambipolar diffusion and magnetic field strengths in \S\ref{sec:pm}. Implications of our results and mechanism of magnetic flux concentration are further discussed in \S\ref{sec:ds}. Finally, we conclude in \S\ref{sec:cc}.

\section{Methods}\label{sec:method}

\subsection{Dynamical Equations}

We use the grid-based high-order Godunov MHD code Athena++ \citep{stone_etal20} to carry out the simulations in this work.  Athena++ is a rewrite from the predecessor Athena code \citep{gs05,gs08,stone_etal08} in C++ language with a comprehensive set of physics and numerical capabilities, and significantly improved performance thanks to its modular design and the use of dynamical execution model.
We solve the equations of non-ideal MHD (as implemented in \citealp{bs17}) using the van Leer time integrator, the HLLD Riemann solver, and the piecewise linear reconstruction. Super time-stepping is employed (based on \citealp{alexiades96}, described in \citealp{simon_etal13b}) to speed up calculations for diffusive non-ideal MHD physics, as was extensively used in our previous works. The dynamical equations in the conservative form read
\begin{gather}
\pdv{\rho}{t} + \div{(\rho \vb{v})} = 0\ , \\
\pdv{(\rho\vb{v})}{t}+\div{ \vb{M}} = -\rho\grad{\Phi}\ , \label{eq:mom}\\
\pdv{\vb{B}}{t}=\curl{(\vb{v}\times\vb{B}}-c\vb{E}^\prime)\ , \label{eq:ind} \\
\pdv{E}{t}+\div{\left[(E+P^*)\vb{v}-\frac{\vb{B}(\vb{B}\cdot\vb{v})}{4\pi}+\vb{S}^\prime\right]} =-\rho(\vb{v}\cdot \grad \Phi) -\Lambda_\mathrm{c}\ . \label{eq:energy}
\end{gather}
Here, $\vb{v}$, $\rho$, and $P$ are gas velocity, density, and pressure, and $\vb{B}$ is the magnetic field vector. The moduli of velocity and magnetic vectors are $v = |\vb{v}|$ and $B = |\vb{B}|$. 
The stress tensor $\vb{M}$ is defined as
\begin{equation}
\vb{M} \equiv \rho\vb{v}\vb{v}-\frac{\vb{B}\vb{B}}{4\pi}+\vb{P^*}\ ,
\label{eq:tensor}
\end{equation}  
where $\vb{P^\ast}=P^\ast\vb{I}$ with $\vb{I}$ being the identity tensor, $P^*=P+P_B$ is the total pressure, and $P_B=B^2/8\pi$ denotes the magnetic pressure. The total energy density is $E = \epsilon+\rho v^2/2+B^2/8\pi$, where $\epsilon$ is the internal energy density and is related to the gas pressure by an ideal gas equation of state $P=({\gamma-1})\epsilon$. We adopt adiabatic index $\gamma=7/5$ for the molecular gas in the outer disks. The cooling term $\Lambda_{\rm c}$ in the last equality will be elaborated in \S\ref{sec:model}. The gravitational potential of the central star is implemented as a source term given by $\Phi=-GM/r$, with stellar mass $M$. 

The components of non-ideal MHD terms manifest in the induction Equation \eqref{eq:ind}. The electric field in the rest fluid frame is 
\begin{equation}
\vb{E}^\prime=\frac{4\pi}{c^2}{(\eta_O\vb{J} +\eta_A\vb{J_\perp})}\ ,
\end{equation}  
where Ohmic and ambipolar diffusivities are denoted by $\eta_O$ and $\eta_A$. The current density is given by $\vb{J}=c\curl\vb{B}/4\pi$, where we express the component of $\vb{J}$ that is perpendicular to the magnetic field by $\vb{J_\perp} = - (\vb{J} \times \vu{b}) \times \vu{b}$, and the unit vector of magnetic field is denoted by $\vu{b}=\vb{B}/B$. The Poynting flux associates to the non-ideal MHD is written as 
\begin{equation}
\vb{S}^\prime=c\vb{E}^\prime\times\vb{B}/4\pi.
\end{equation}  
The above governing equations are written in Gaussian units, whereas in code units the $4\pi$ factors are absorbed to the definition of magnetic pressure such that the magnetic permeability $\mu_0=1$. 

Our simulations are conducted in spherical-polar coordinates $(r,\theta,\phi)$. In code units,
we choose $GM=R_0=1$, with $R_0$ a reference radius at the inner radial boundary. The domain spans over $r\in[1,100]$ in radius, in polar angles over $\theta\in[0,\pi]$, and in azimuthal angles over $\phi\in[0,\pi/4]$. The radial and theta domain encompasses adequate dynamical range to fully accommodate the launching of magnetized disk wind. We also consider cylindrical coordinates, with $R=r\sin\theta$ and $z=r\cos\theta$, to better describe simulation setup and diagnostics. 

\subsection{Disk Model}\label{sec:model}

The disk model is scale-free and is largely identical to that of \citet{bs17} and \citet{cb20}. The density profile comprises of separate radial and theta components
\begin{equation} 
\rho(r, \theta) = \rho_0\left(\frac{r}{R_0}\right)^{-q_D}f(\theta)\ ,  \label{eq:den}
\end{equation}
where $\rho_0$ is the midplane density at $r=R_0$, and we set $q_D=2$. The function $f(\theta)$ is computed by solving hydrostatic equilibrium based on a prescribed equilibrium temperature profile $T_{\rm eq}$, which is specified as
\begin{equation}
T_{\rm eq}(r, \theta)\equiv\frac{P}{\rho}=T_0\frac{R_0}{r}\epsilon^2(\theta)\ , \label{eq:temp}
\end{equation}
where $T_0=v_{\rm K0}^2=GM/R_0$ is the midplane temperature at $r=R_0$. The aspect ratio is $\epsilon(\theta)= H/r$, where $H=c_s/\Omega$ is the pressure scale height, and $c_s^2 =P/\rho$ is the isothermal sound speed. The relation between $f(\theta)$ and $\epsilon(\theta)$ can be found in \citet{bs17,cb20} by solving hydrostatic equilibrium. We choose the temperature profile $\epsilon(\theta)$ to be
\begin{gather}
\epsilon(\theta) = \epsilon_{\mathrm{d}} + \frac{1}{2}(\epsilon_{\mathrm{w}} - \epsilon_{\mathrm{d}})\left[\tanh\left(\frac{\delta \theta}{\mathrm{\epsilon_d/2}}\right)+1\right]\ ,
\end{gather}
which transitions from a disk value $\epsilon_\mathrm{d} = 0.1$ to a surface value $\epsilon_{\rm w} =0.5$ smoothly. We thus quote $H_d=0.1R$ as the disk scale height. The transition peaks at $\theta_\mathrm{trans} = 3.5H_d/r$ and is mostly achieved within a width of $0.05$ radian, with $\delta \theta = |\theta - \pi/2|- \theta_\mathrm{trans}$ the angle that deviates from $\theta_\mathrm{trans}$. Such choice mimics additional heating from UV and X-rays in the disk atmosphere (e.g., \citealp{glassgold_etal04,walsh_etal10}). The resulting equilibrium rotation profile is given by
\begin{equation}
v_\phi^2(r,\theta)=\frac{GM}{r}-(q_D+1)T_0R_0\frac{\epsilon^2(\theta)}{r}\ ,
\end{equation}
while radial and meridional velocities are zero at equilibrium. On the other hand, to trigger the MRI, we apply random noise to these velocity components with an amplitude of $5\%$ of the local sound speed.

We initialize magnetic fields only for poloidal component by prescribing an azimuthal vector potential \citep{zanni_etal07}:
\begin{equation}
A_\phi(r,\theta) = \frac{2B_{z0}R_0}{3-q_D}\left(\frac{R}{R_0}\right)^\frac{1-q_D}{2}[1+(m\tan \theta)^{-2}]^{-\frac{5}{8}}\ .
\label{eq:Aphi}
\end{equation}
The poloidal field is computed by $\vb{B}=\nabla \times \vb{A_\phi}$, allowing preservation of divergence-free condition.
In the above, $B_{z0}$ is the midplane field strength at $R_0$. The parameter $m$ reflects the degree to which initial poloidal field lines bend.
We choose $m=0.5$, corresponding to modestly bent initial fields. At the midplane, we obtain a pure vertical field
\begin{equation}
B_{\mathrm{mid},z}(R)=B_{z0}\bigg(\frac{R}{R_0}\bigg)^{-(q_D+1)/2}\ .
\end{equation}
The strength of the poloidal fields is characterized by plasma $\beta$, the ratio of gas to magnetic pressure, which is constant at the disk midplane in our prescription. We choose a midplane value of $\beta_0=10^4$ in most of our simulation models.

We relax disk temperature to initial temperature through the $\mathrm{\Lambda_c}$ term on the right hand side of Equation \eqref{eq:energy}. After each time step over a time interval $\Delta t$, we adjust gas temperature $T$ to $T+\Delta t$ with
\begin{equation}
\Delta T =(T_\mathrm{eq}-T)[1-\exp(-\frac{\Delta t}{\tau})] \,
\end{equation}
where the relaxation time $\tau$ is prescribed as a fraction of the local Keplerian orbital period, $\tau(R)\propto \mathrm{P_{orb}}\equiv 2\pi/\Omega_\textrm{K}(R)$. To avoid the excitation of potential hydrodynamic instabilities in the disk, primarily vertical shear instability (VSI), we choose $\tau= \mathrm{P_{orb}}$ in the disk region. We further make $\tau$ smoothly transition to 0 in the wind zone (i.e., locally isothermal) by strictly setting to equilibrium values at each time step. This allows the system to better relax to a quasi-steady state based on our numerical experiments.

\subsection{Els\"{a}sser Numbers}\label{ssec:elsasser}

The dimensionless Els\"{a}sser numbers quantify the strength of non-ideal MHD effects. For AD, it is given by
\begin{equation}
{\rm Am}=\frac{v_\textrm{A}^2}{\eta_\textrm{A}\Omega_\textrm{K}}\ ,
\end{equation}
where $v_\mathrm{A}=\sqrt{B^2/4\pi \rho}$ is the Alfv\'{e}n speed. Given an ionization fraction,
ambipolar diffusivity typically scales as $\eta_A\propto B^2/\rho^2$,
and hence {\rm Am} does not depend on the field strength. 
In our simulations, {\rm Am} is set to a constant throughout the bulk disk (Table \ref{table:runs}), and transitions to ${\rm Am}=100$ from the disk zone towards the wind zone to mimic the elevated ionization level above the disk surface mainly by stellar X-rays and far-UV (FUV, \citealp{pc11}). The transition of {\rm Am} follows the same form as that of temperature, since they share the same physical origins. Correspondingly, we consider the transition height, characterized by $\theta_{\rm trans}=0.35$, to be the ``wind base", as identified in earlier studies \citep{bs13,gressel_etal15}.

Besides AD, we also incorporate Ohmic resistivity near the inner boundary for pure numerical considerations.
The resulting magnetic dissipation allows poloidal magnetic fields to stay straight, which helps stabilize the inner boundary. The value of midplane resistivity is fixed to $\eta_\mathrm{O} = 0.02c_sH$ or $0.03c_sH$ at $R=R_0$ and is gradually relaxed in radial direction to approach approximately zero at $R=2$. In $\theta$ direction, $\eta_\mathrm{O}$ transitions from the midplane value to zero around $1.2$ radian above/below midplane (near the pole) over a thickness of $\sim H_d/r$, and we strictly require both AD and resistivity to vanish at two active cells from the pole. 

The Hall effect can be marginally important at the midplane region up to $\sim 30-60$ AU \citep{bai15}.
%However, it has only driven weak turbulence across the disk midplane as in local shearing box simulations. 
Nevertheless, it was found in local simulations to have only minor impact on angular momentum transport. 
Hence, as a first study, we do not include the Hall effect to avoid potential complications.

\subsection{The Mesh and SMR}\label{sec:SMR}

\begin{figure*}
    \centering
    \begin{multicols}{1}
    \begin{center}
        {\includegraphics[width=0.25\textwidth]{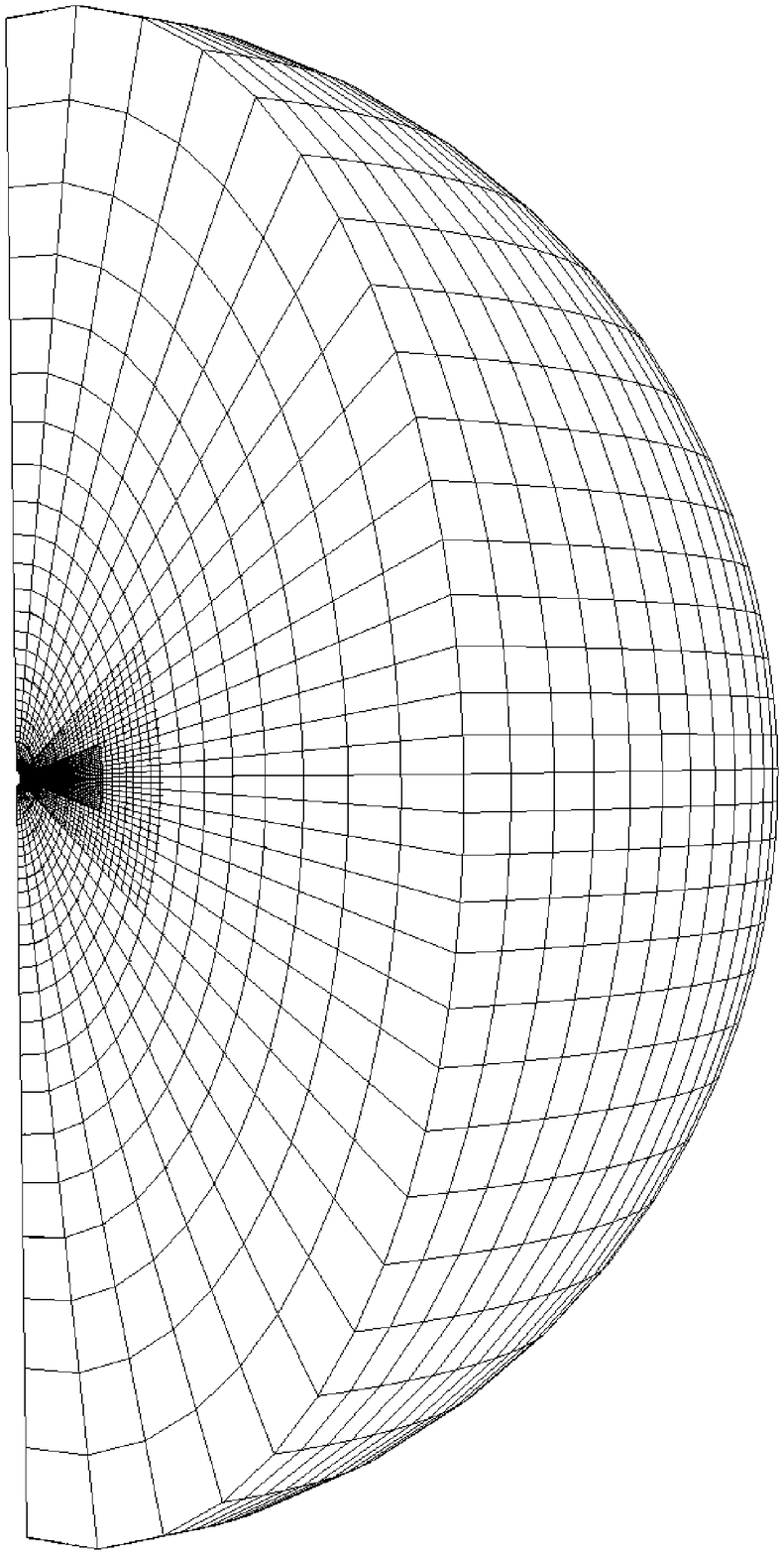}}\par 
        {\includegraphics[width=0.4\textwidth]{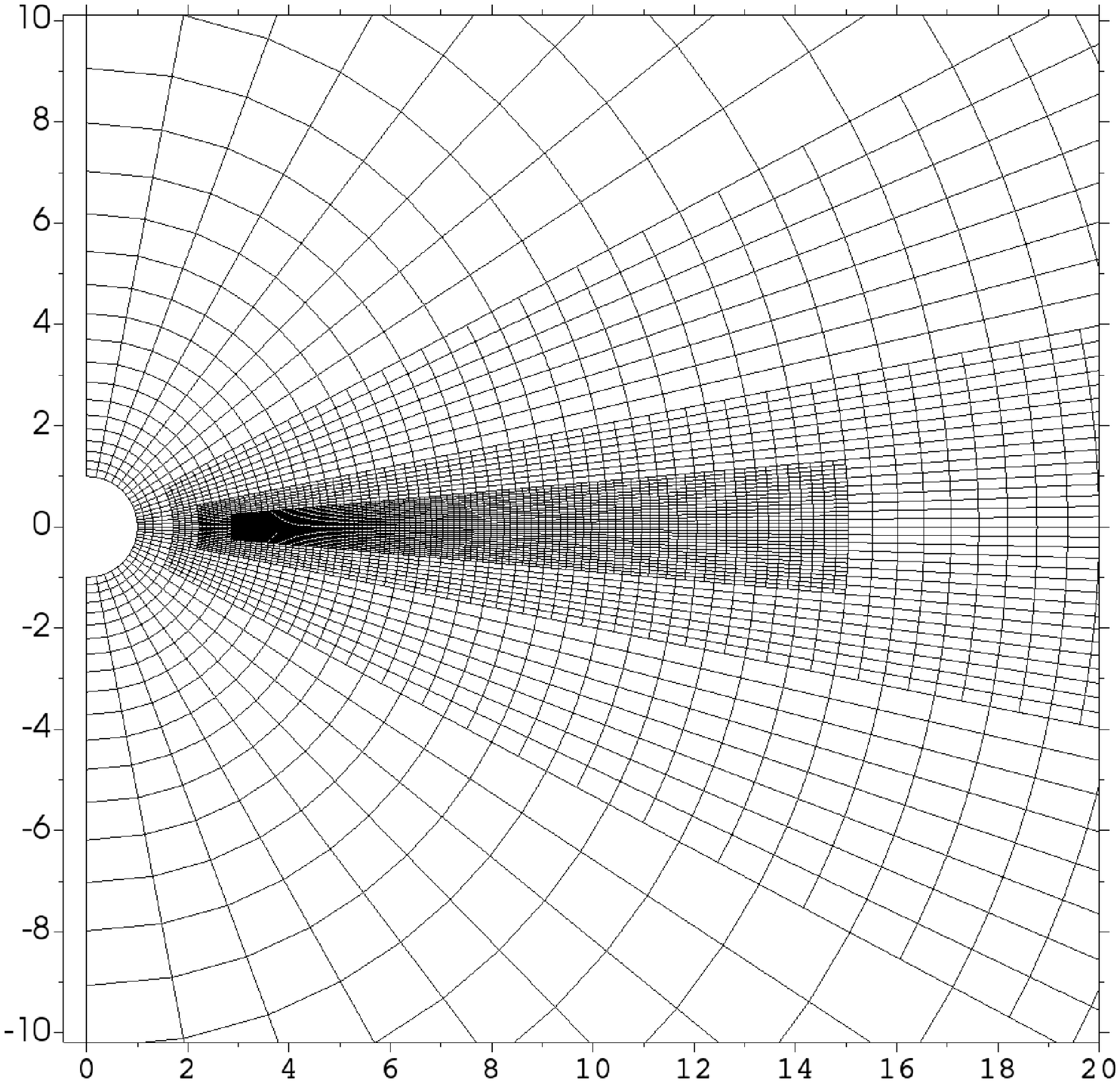}}\par 
    \end{center}
    \end{multicols}
\caption{Left: the 3D mesh that covers the entire simulation domain. Right: a zoom-in mesh in $R,z$-plane of inner disks. The meshes in both panels are plotted by every 4 cells in radial direction and every 2 cells in $\theta$ direction for illustrative purposes. 
Note that due to visualization artifacts, straight line segments in $\theta$ and $\phi$ should be considered as arcs.} 
\label{fig:mesh}
\end{figure*}

In spherical-polar coordinates, the most severe limitation to simulation timestep is the small azimuthal grid size in the polar region. Taking the advantage of hierarchical mesh structure in Athena++, we use static mesh refinement (SMR), which allows us to use coarse grid in the polar region, and in the meantime to zoom in the disk zone to achieve adequate resolution.

All 3D simulations carried out in this paper are equipped with the same grid structure. At the root level, we set $144\times 64 \times 12$ active cells in $(r,\theta, \phi)$. The radial domain spans from $r=1$ to $r=100$ with logarithmic grid spacing. The grid spacing in $\theta$-direction increases by a constant factor of 1.05 per root grid cell from the midplane towards the two poles, yielding
a resolution at the midplane about five times finer than that at the pole. 
The grid spacing in the $\phi$ direction is uniform. 

On top of the root level, we set three levels of refinement. The mesh structure is shown in Figure \ref{fig:mesh}. The left panel illustrates the mesh of the entire simulation domain, and the right panel zooms in to the inner parts of the disk. In the radial direction, the finest level spans over $r\sim3-15$, where our data analysis are conducted within this range. In $\theta$-direction,
the finest level $n=3$ covers $\pm H$, the $n=2$ level occupies $\pm 2.5H$, $n=1$ level fills $\pm 6H$.
Overall, this grid setting enables us to achieve a resolution of 32 cells per $H$ in $r$, 30 cells per $H$ in $\theta$ at the disk midplane, and 12 cells per $H$ in $\phi$ in the $n=3$ level.

\subsection{Boundary Conditions}

We choose a ``fixed-state" type of inner boundary condition,
which we find is able to sufficiently stabilize the inner boundary so that it does not strongly interfere with the bulk computational domain (see further discussions in Section \ref{ssec:num}).
Specifically, we impose temperature to be equal to their equilibrium values from Equation \eqref{eq:temp}. Density is extrapolated from the last active zone, assuming $\rho\propto r^{-qD}$.
The angular velocity is set to the minimum between equilibrium $v_\phi$ and the angular velocity of solid body rotation $\Omega_\mathrm{K}(R_0)R$. The other two velocity components are $v_r=v_\theta=0$. 
The addition of resistivity mentioned in Section \ref{ssec:elsasser} further helps stabilize the inner boundary by creating less turbulent environments and straightening field lines.  
In some simulations which yield strong turbulence, we further enforce the poloidal magnetic field lines penetrating through the inner boundary to be tied to its initial position (no sliding) by resetting the $\phi-$ electric field $E_\phi=0$ at the inner boundary.

Other boundary conditions are standard. At outer radius, the hydrodynamic variables are extrapolated from the last active zone, assuming $\rho \propto r^{-q_D}$, $T \propto r^{-q_T}$ and $v_\phi \propto r^{-1/2}$. Radial and meridional velocities $v_r$ and $v_\theta$ are copied directly from the last active zone, except setting $v_r = 0$ when $v_r<0$ (standard outflow boundary condition). Magnetic field variables of the inner and outer ghost cells are set based by $B_r \propto r^{-2}$, $B_\theta \propto \mathrm{const.}$, $B_\phi \propto r^{-1}$. 
Our $\theta$-domain directly reaches the poles, and we employ polar wedge boundary conditions \citep{zhu18}. Lastly, boundary condition in $\phi$-direction is periodic. 

\subsection{Simulation Models}

Table \ref{table:runs} lists our simulations models. Model names with an `E' indicates those we enforce $E_\phi=0$ at the inner boundary.
In this work, two physical parameters are under investigation, the ambipolar Els\"{a}sser number $\rm Am$ and disk magnetization $\beta_0$. The parameter spaces cover $\beta_0\in\{10^3, 10^4\}$ and ${\rm Am}\in\{0.1, 0.3, 1, 3,10\}$. In addition, all simulations are run up to $3000P_0$, where $P_0$ is the orbital period at the inner boundary $R_0=1$.

We choose fiducial model to be \textsc{\textsc{Am1}} with $\rm Am=1$ and $\beta_0=10^4$. This is due to the fact that the Ambipolar Els\"{a}sser number is found to be on the order of unity towards outer PPD regions \citep{bai11a,bai11b}, and controls the onset of MRI with a threshold $\sim 1$ (e.g., \citealp{bs11}). Moreover, the strength of disk magnetization that yields an accretion rate similar to the typically observed values is on the order of $\beta_0=10^4$ \citep{simon_etal13a,bai15}.

\begin{table}
 \caption{List of Simulation Models and Parameters. } \label{tab:anysymbols}
 \begin{tabular*}{0.9\columnwidth}{@{}l@{\hspace*{70pt}}c@{\hspace*{70pt}}r@{}}
  \hline
  Run & $\beta_0$ & Am\\
  \hline
  \textsc{\textsc{Am1}} & $10^4$  & 1  \\
  \textsc{B3} & $10^3$ & 1 \\
  \textsc{Am0.1} & $10^4 $ & 0.1 \\
  \textsc{Am0.3} & $10^4$  & 0.3  \\ 
  \textsc{Am3E} & $10^4$  & 3 \\
  \textsc{\textsc{Am1}0E} & $10^4$  &  10 \\
  \hline
 \end{tabular*}
\label{table:runs} 
\end{table}

\section{Diagnostics}\label{sec:diag}

\subsection{Mass Accretion}

In this section, we present some relevant diagnostics that will assist data analyses in result sections. 
In cylindrical coordinates, the $\phi-$component of momentum equation \eqref{eq:mom} can be cast to
expresses the conservation of angular momentum:
\begin{equation}
\begin{split}\label{eq:am}
&\pdv{(2\pi R \Sigma j)}{t}  + \pdv{(\dot M_{\rm acc} j)}{R}   \\ 
&+ \pdv{}{R}\bigg[2\pi R^2 \int^{z^+}_{z_-}   \langle T_{R\phi} \rangle \; dZ \bigg]
+2\pi R^2  \; \langle T_{z\phi} \rangle  \big|^{z^+}_{z_-}  = 0\ ,
\end{split}
\end{equation}
where $j=\langle v_\phi R\rangle\approx v_{\rm K} R$ is the specific angular momentum,
$\Sigma =\int^{z^+}_{z_-} \rho dz$ is the disk surface density,
angle brackets $\langle \cdot \rangle$ denote azimuthal and temporal averages. We truncate the disk at $z^\pm=\pm0.32R$, which is slightly smaller than the expected position of $\tan(0.35)R$. This is chosen because we impose the transition over a finite thickness (instead of a sharp jump) of $\sim\epsilon_d/2=0.05$ in $\theta$ and we do observe that wind can be launched at a lower height.

The accretion rate is defined as
\begin{equation}
\dot{M}_\mathrm{acc} = -2\pi R  \int^{z^+}_{z_-}   \rho v_R dz\ , 
\label{eq:acc}
\end{equation}
and the $R\phi$ and $z\phi$ components of the stress tensor are given by
\begin{equation}
T_{R\phi} =  \rho\delta v_R\delta v_\phi - \frac{B_RB_\phi}{4\pi}\ ,\quad
T_{z\phi}=  \rho v_zv_\phi- \frac{B_z B_\phi}{4\pi}\ ,
\end{equation}
where $\delta v_R$ and $\delta v_\phi$ are the radial and azimuthal velocity fluctuations.
The hydrodynamic and magnetic parts in the above are known as the Reynolds and
Maxwell stresses, respectively, which we denote as $T^{\rm Rey}$ and $T^{\rm Max}$.

In Equation \eqref{eq:am}, the first term corresponds to the change of angular
momentum due to the time evolution of surface density, the second term is the
advection of angular momentum from the accretion flow, the third and last terms
are the driving forces of disk angular momentum transport, giving the fluxes of
angular momentum in the radial and vertical dimensions, respectively.

We note that from the continuity equation, we have
\begin{equation}\label{eq:acc_cont}
2\pi R\pdv{\Sigma}{t}=\pdv{\dot{M}_{\rm loss}}{R}-\pdv{\dot{M}_{\rm acc}}{R}\ ,
\end{equation}
where $\dot{M}_{\rm loss}(R)$ is the cumulative disk mass loss rate within radius $R$, and it
can be more conveniently expressed as
\begin{equation}
    \frac{d\dot{M}_{\rm loss}}{d\log R}=2\pi R^2\rho v_z\bigg|^{z^+}_{z^-}\ ,
\end{equation}
where note that we use $\log{R}$ to ensure the dimension in the above equation to match
that of the accretion rate.
By substituting (\ref{eq:acc_cont}) to Equation \eqref{eq:am}, we obtain
\begin{equation}
\begin{split}\label{eq:am1}
&2\pi R \Sigma\pdv{j}{t}  + \dot M_{\rm acc}\pdv{j}{R}   \\ 
&+ \pdv{}{R}\bigg[2\pi R^2 \int^{z^+}_{z_-}   \langle T_{R\phi} \rangle \; dz \bigg]
+2\pi R^2  \; \langle T_{z\phi}^{\rm Max} \rangle  \big|^{z^+}_{z_-}  = 0\ .
\end{split}
\end{equation}
Here, only the Maxwell stress is retained for vertical transport of angular momentum. 

In quasi-steady state, we anticipate the specific angular momentum $j\sim v_KR$
to be stationary in space and time, with ${\partial j}/{\partial R}=v_K/2$. Therefore, in our standard analysis, we divide the above equation by $v_K/2$, so that the second term becomes $\dot{M}_{\rm acc}$ if $j=v_KR$. This approach, which we refer to as ``original" decomposition, directly illustrates the expected contribution to accretion from viscous stress (3rd term) and wind (last term). There is also the first term, which reflects the time-variation in specific angular momentum. We call it time-dependent correction, and it can arise when disk forms ring-like substructures which modifies the rotation profile.

In the meantime, as the formation of disk substructure makes rotation profile deviate from Keplerian, implies that $\partial j/\partial R$ does not necessarily be equal to $v_K/2$. To obtain the actual contribution from these physical process to $\dot{M}_{\rm acc}$, one needs to, on top of the ``original" decomposition, multiply this equation by a dimensionless correction factor ${\rm Corr}(R)\equiv(v_K/2)(\partial j/\partial R)^{-1}$. We call this approach ``actual" decomposition.

The radial transport of angular momentum is mainly mediated by turbulence in the disk, and
is characterized by the classic dimensionless $\alpha$ parameter \citep{ss73},  
\begin{equation}
\alpha =T_{R\phi}/P\ .
\label{eq:alp}
\end{equation}
Over the disk column, the vertically averaged $\alpha$ is obtained by
\begin{equation}
\alpha=\frac{\int^{z^+}_{z_-} T_{R\phi} \ dz}{\int^{z^+}_{z_-} P \ dz}\ .
\end{equation}

The vertical transport of angular momentum is mediated by the magnetized disk winds. 
Assuming magnetized winds drive disk accretion in steady state with $j=v_KR$, the local mass flux can
be written by \citep{wardle07,bs13} 
\begin{equation}
\frac{1}{2}\OmK \rho v_R \approx \frac{B_z}{4\pi} \pdv{B_\phi}{z},
\label{eq:mf}
\end{equation}
where fast accretion occurs at steep vertical gradient of $B_\phi$.
This relation is derived in laminar state, while we will examine our results under turbulent
conditions by taking averages using $\langle\rho v_R\rangle$, $\langle B_z\rangle$ and $\langle B_\phi\rangle$.

\subsection{Units Conversion}\label{ssec:uc}

Since the code units are scale-free, here we present the conversion of code units to physical units for accretion rates $\dot{M}$ and magnetic field strength $B$. Given a radius in code units, we aim to find the corresponding mass accretion rate and magnetic field strength in realistic PPDs.
Consider at a physical radius of $30$ AU, the surface density is $10$ g cm$^{-2}$ as for minimum-mass solar nebula \citep{weidenschilling77}. 
The mass accretion rate in physical units is expressed by
\begin{equation}
\begin{split}\label{eq:mdot_conv}
    &\dot{M}_{\rm phys} \approx 1.55\times10^{-8}M_{\odot}\ {\rm yr}^{-1}\bigg(\frac{\dot{M}_{\rm code}}{10^{-4}}\bigg)\bigg(\frac{M}{M_\odot}\bigg)^{1/2}\\
    &\quad \times\bigg(\frac{R_{\rm phys}}{\rm 30AU}\bigg)^{1/2}\bigg(\frac{R_{\rm code}}{R_0}\bigg)^{q_D-3/2}\bigg(\frac{\Sigma}{10{\rm g}\ {\rm cm}^{-2}}\bigg) \ ,
\end{split}
\end{equation}
and the magnetic field strength is given by
\begin{equation}
\begin{split}\label{eq:B_conv}
\frac{B_{\rm phys}}{B_{\rm code}} \approx 0.81{\rm G}&\ \bigg(\frac{R_{\rm phys}}{30 \rm AU}\bigg)^{-1} \bigg(\frac{R_{\rm code}}{R_0}\bigg)^{(q_D+1)/2}\\ &\times\bigg(\frac{M}{M_\odot}\bigg)^{1/2}\bigg(\frac{\Sigma}{10 {\rm g\ cm^{-2}}}\bigg)^{1/2}\ .
\end{split}
\end{equation}
In the above, subscripts ``${\rm phys}$" and ``${\rm code}$" represent quantities measured in physical units and code units, respectively. These results will be utilized later for discussions.

\subsection{Characterizing Rings and Gaps}

Later in \S\ref{sec:fc} and \S\ref{sec:pm}, we will see
the disk form annular substructures with variations in surface density profile. For the ease of analysis, we define the amplitude of surface density variation (depth $\delta_\Sigma$) of a ring and a gap as 
\begin{equation}
\delta_\Sigma \equiv \frac{\Sigma_{\rm max}-\Sigma_{\rm min}}{\Sigma_{\rm min}},
\label{eq:sdv}
\end{equation}
which is the difference between the local maximum and minimum of surface density divided by the minimum.
The width $w_\Sigma$ of a ring or gap is measured as the half distance of two adjacent local maxima in surface density $\Sigma$,
\begin{equation}
w_\Sigma \equiv \frac{r_{\rm out, max}-r_{\rm in, max}}{2}. 
\end{equation}
Note that our definition of the depth and width are purely based on gas surface density, which can be directly obtained from simulation data. In observations, since the surface density is not directly observable, the gap depth is usually defined as the difference between the intensity at the gap location and the peak value of its outer ring, and the gap width is as the full width at half depth (e.g. \citealp{long_etal18}). The definitions of gap depth and width in surface density also vary in the literature (see Section 3.2 in \citealp{df17}). In this paper, we employ simple definitions of the depth and width in order to give a semi-quantitative description of our simulation results instead of directly comparing with observed characteristics of rings and gaps.

\section{The fiducial model}\label{sec:fi}

\begin{figure*}
\includegraphics[width=1\textwidth]{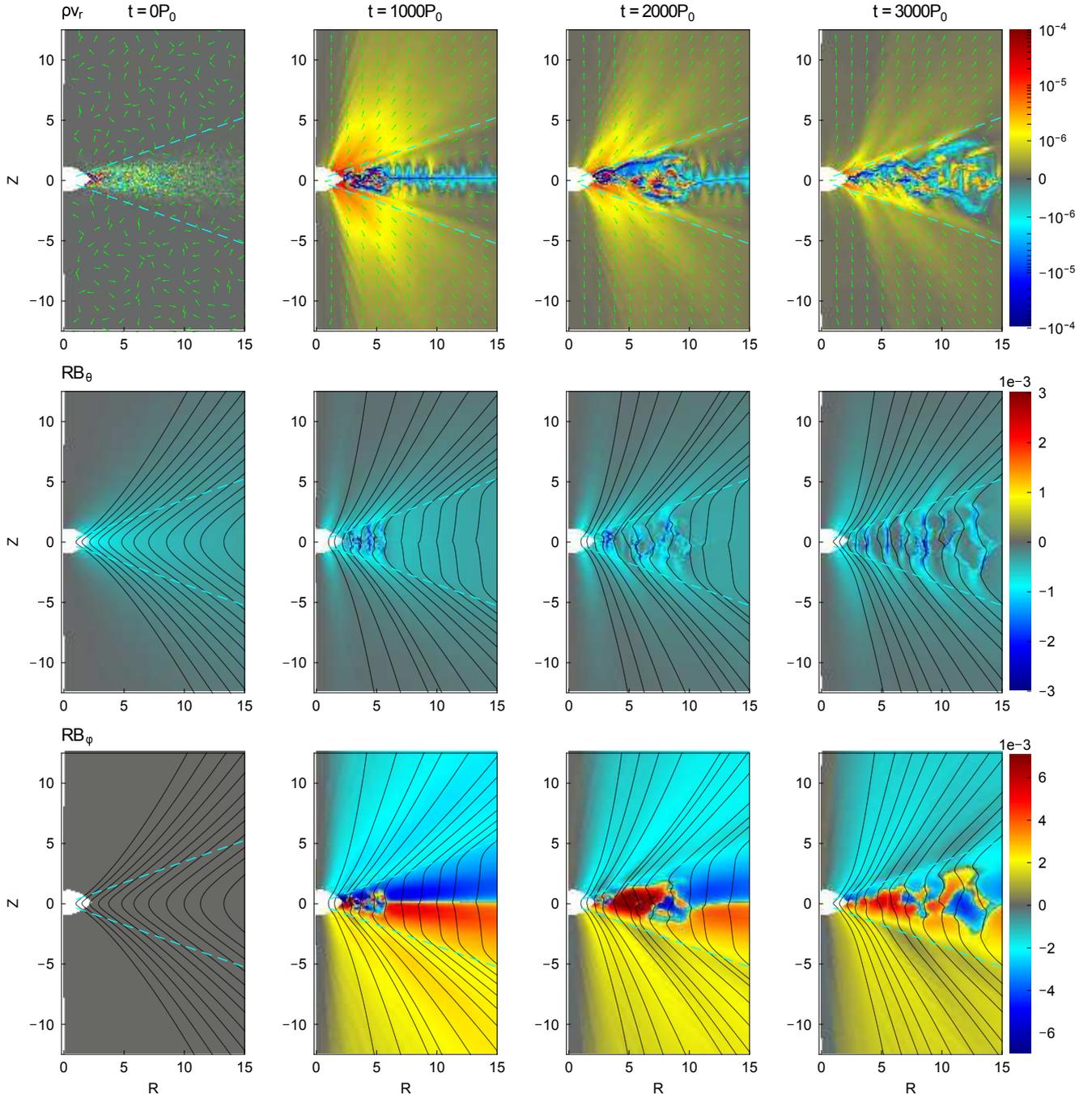}
\caption{
Snapshots of azimuthally averaged radial mass flux and magnetic fields for Model $\textsc{Am1}$ at $t = 0, 1000, 2000, 3000P_0$ from left to right. Top panels: snapshots of radial mass flux $\rho v_r$, overlaid with green arrows of velocity vectors. Middle panels: snapshots of scaled meridional magnetic field $RB_\theta$. Bottom panels: snapshots of scaled toroidal magnetic field $RB_\phi$. Overlaid black curves are equally spaced contours of poloidal magnetic flux. Dashed curves mark an opening angle of $0.35$ above and below the midplane, which correspond to the transition from non-ideal MHD dominated disk zone to ideal MHD dominated wind zone.}
\label{fig:ct}
\end{figure*}

In this section, we focus on the fiducial model \textsc{\textsc{Am1}}, following by detailed analyses on the overall disk evolution (\S\ref{sec:oe}), magnetic flux concentration (\S\ref{sec:fc}), magnetic flux transport (\S\ref{sec:ft}), turbulence and stress levels (\S\ref{sec:tu}), angular momentum transport (\S\ref{sec:am}), and azimuthal features (\S\ref{sec:af}). We present time in units of the innermost orbital time $P_0\equiv P_{\rm orb} (R_0) =2\pi$ throughout the result sections.

Based upon the above discussion, we choose to analyze data over a time domain $t\in [2600, 3000]P_0$ and focus on a spatial domain $R \in [5, 10]$ if not otherwise noted. We quote that $2600P_0$ is approximately $233$ and $82$ local orbits at $R=5$ and $R=10$, respectively, and $3000P_0$ corresponds to $268$ and $95$ local orbits. The MRI is well developed to a radius of $R\sim 15$ over the time interval selected. 

\subsection{Overall Disk Evolution}\label{sec:oe}

The time evolution of the fiducial Model \textsc{Am1} is illustrated in Figure \ref{fig:ct}. We show snapshots of radial mass fluxes (top), poloidal (middle) and toroidal (bottom) magnetic field strength at different innermost orbital times $t=0,1000,2000,3000P_0$. The overlaid %approximately vertical white contour lines delineate linearly 
black contours represent equally spaced lines of constant magnetic flux $\Phi$, which represents the cumulative magnetic flux from the north pole, calculated by 
\begin{equation}
\Phi(r,\theta)= \int_0^\theta \ \overline{B}_r(r,\theta) r^2 \sin\theta \ d\theta\ ,
\end{equation}
or equivalently, by the divergence free condition $\nabla\cdot\mathbf{B}=0$,
\begin{equation}
\Phi(r,\theta)=\Phi(r_0,\theta)+\int_0^r \ \overline{B}_\theta(r,\theta)\ r\sin\theta \ dr\ ,
\end{equation}
where $\overline{B}_r$ and $\overline{B}_\theta$ are azimuthally averaged, and $\Phi(r_0,\theta)$ is the magnetic flux at the inner radial boundary. 

At $t=0P_0$, we see the smooth profile of poloidal field imposed throughout the disk on top of initial velocity perturbations. The large-scale poloidal field threading the disk results in launching of magnetized winds from surface layers. The Keplerian rotation winds up the poloidal field to generate the toroidal magnetic field, and it dominates the magnetic field strength over the entire course of the simulation. At $t=1000P_0$, for a large extent of the disk ($R \gtrsim 6$), toroidal fields are symmetric but possess opposite signs about the midplane, generating a thin current layer there and driving the gas accretion. The accretion flow can be clearly seen in the top panel for radial mass flux, where efficient inward transport of mass is discernible at the midplane, consistent with previous global 2D simulations (e.g. \citealp{suriano_etal18,bs17,gressel_etal20}). The MRI turbulence sets in progressively from inner regions towards larger radii. At the inner part of the disk ($R\lesssim 6$), the MRI turbulence leads to the disruption of current sheet at the midplane. 

As the system evolves, at $t=2000P_0$ and $t=3000P_0$, the disk becomes fully MRI turbulent up to a radius of $R\sim 10$ and $15$, as seen from the disruption of the midplane current sheet. Field strength is still dominated by the toroidal component in the bulk disk. As the mean toroidal fields flip at arbitrary vertical locations, the resulting accretion flow becomes randomly distributed in the disk. We also see that the distribution of poloidal magnetic flux within the disk becomes non-smooth, as opposed to the initial smooth profile. Despite the disk is fully turbulent, the field configuration in the wind region remains smooth.

\begin{figure*}
\centering
\includegraphics[width=1\textwidth]{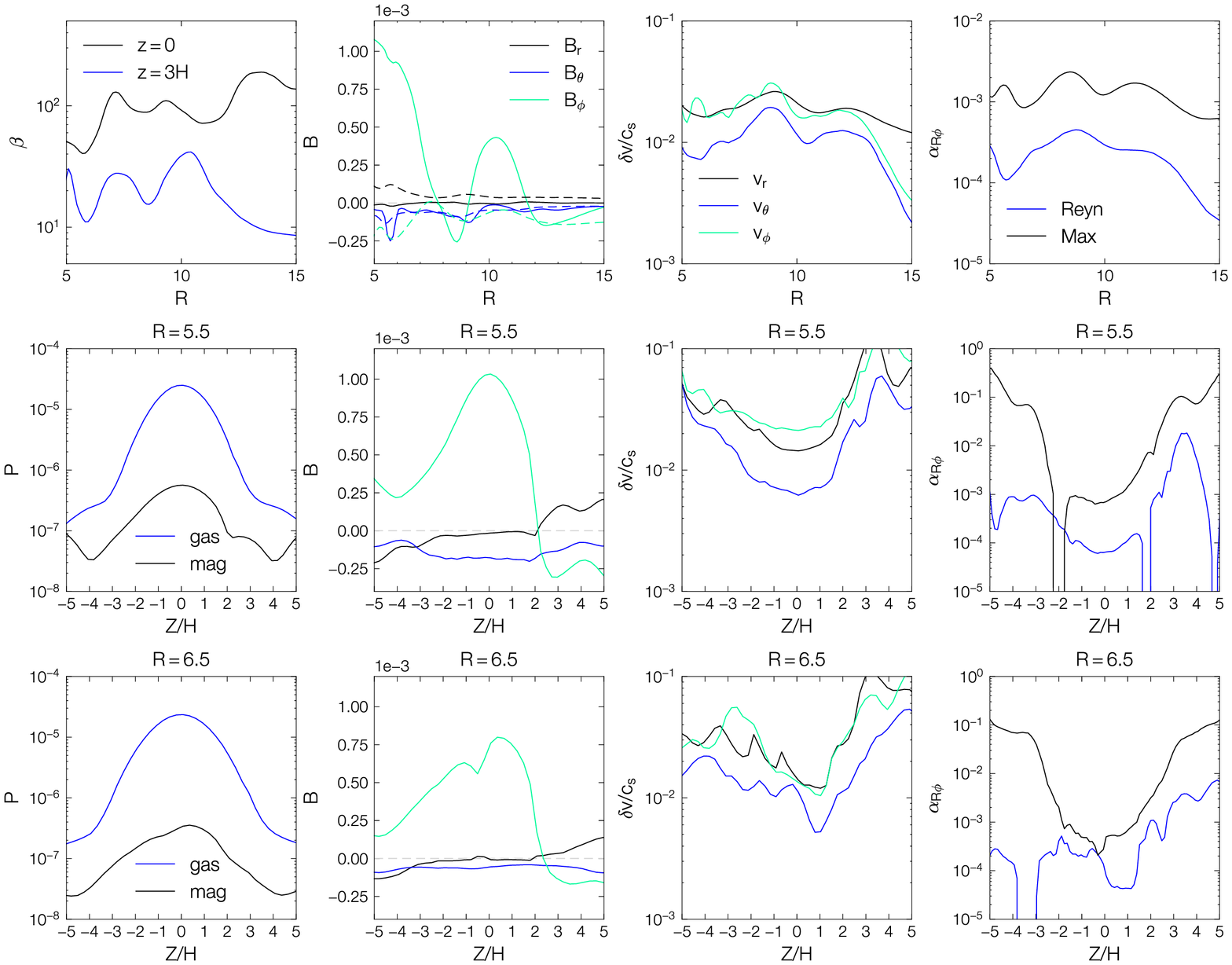}
\caption{First column: radial profiles of plasma $\beta$ at $z=0$ and $z=3H$ (top); vertical profiles of gas pressure $P$ and magnetic pressure $P_B$ at $R=5.5$ (middle) and $R=6.5$ (bottom).
Second column: radial profiles of three components of magnetic field at $z=0$ (solid) and $z=3H$ (dashed)  (top); vertical profiles of three components of magnetic field at $R=5.5$ (middle) and $R=6.5$ (right). 
Third column: three components of velocity fluctuations in units of the local sound speed $c_s$. 
Forth column: dimensionless $\alpha$ parameter of Reynolds and Maxwell stresses.
}
\label{fig:fig8}
\end{figure*}

In the left two columns of Figure \ref{fig:fig8}, we show diagnostic quantities at the end of simulation, averaged over the last $400P_0$, including plasma $\beta$ at $z=0$ and $z=3H$, gas and magnetic pressure, and three components of magnetic field strength. We also note that the MRI is mainly driven by the vertical field, as found in previous local simulations (e.g., \citealp{bs11,simon_etal13a,simon_etal13b}). These radial and vertical profiles will be discussed in the subsequent subsections to facilitate the analysis. 

\subsection{Magnetic Flux Concentration and Evolution}\label{sec:fc}

\begin{figure*}
\centering
\includegraphics[width=0.9\textwidth]{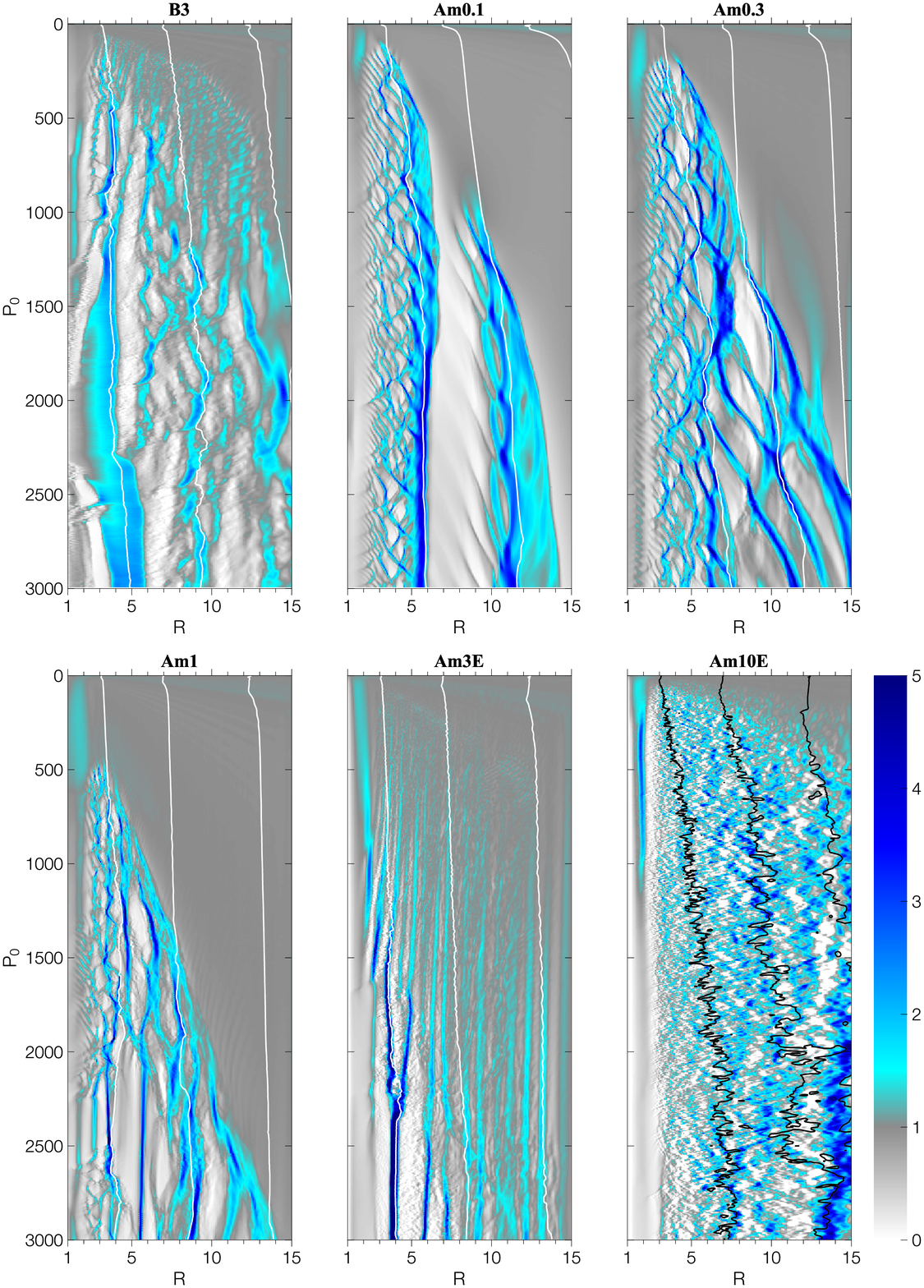}
\caption{Space-time plot of azimuthally and vertically ($\pm1 H$) averaged midplane $B_z/B_z(t = 0)$ for all simulation models. Overlaid white or black vertical curves delineate linearly equally spaced contour lines of magnetic flux $\Phi$.}
\label{fig:st}
\end{figure*}

\begin{figure*}
\includegraphics[width=0.9\textwidth]{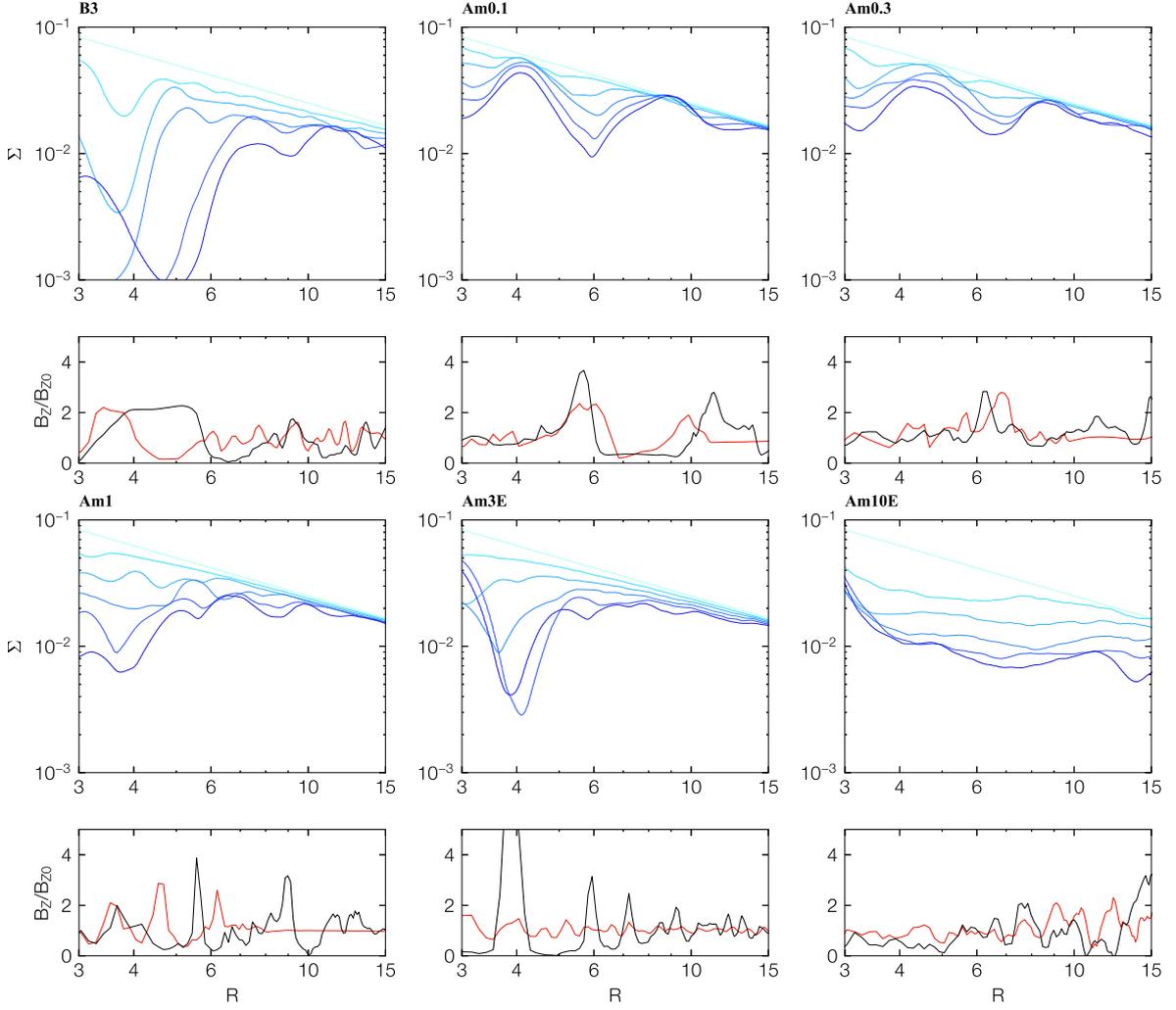}
\caption{Surface densities $\Sigma$ (first row and third row) as a function of radius at $t=0,600,1200,1800,2400,3000P_0$ from light to dark colors of all simulation models. Vertical magnetic fields normalized by initial values $B_Z/B_{Z0}$ (second row and forth row) as a function of radius at $t=1200P_0$ (red) and $t=3000P_0$ (black).}
\label{fig:Sig}
\end{figure*}

\begin{figure}
\includegraphics[width=0.48\textwidth]{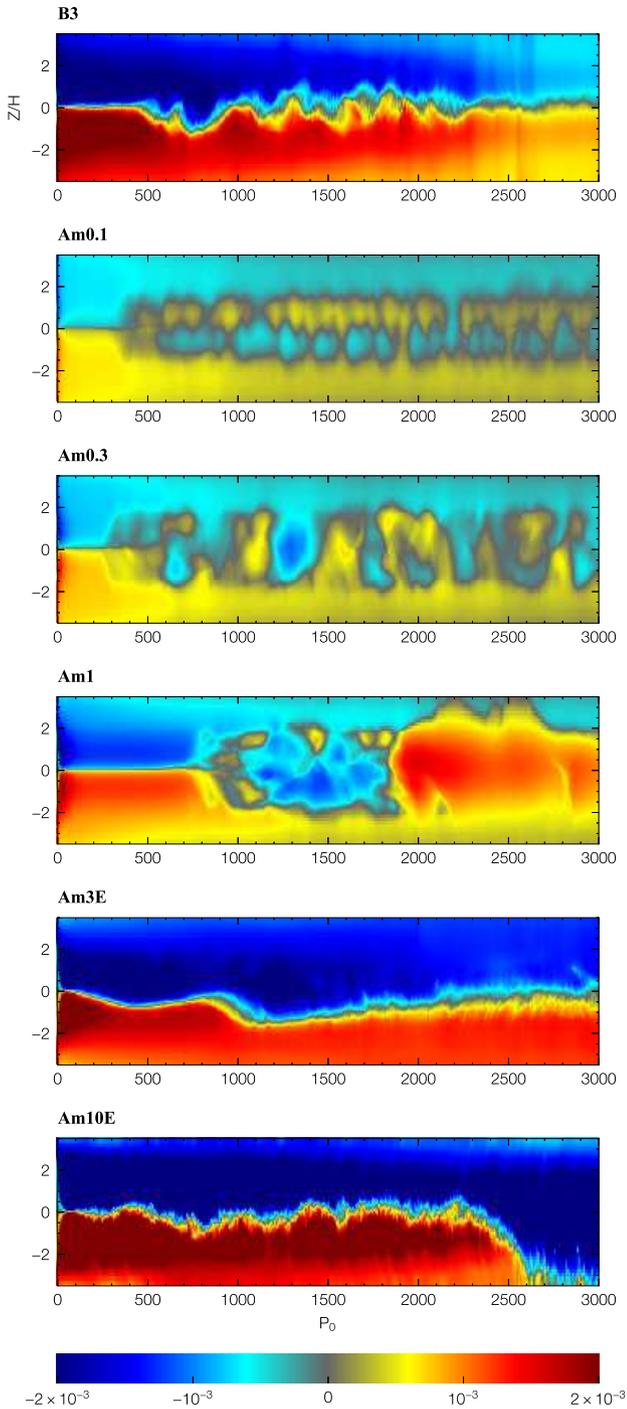}
\caption{The space-time diagrams of azimuthally-averaged toroidal magnetic field $B_\phi$ at $r=5.5$ for all simulation models.}
\label{fig:DY}
\end{figure}

\begin{figure}
\centering
\includegraphics[width=0.45\textwidth]{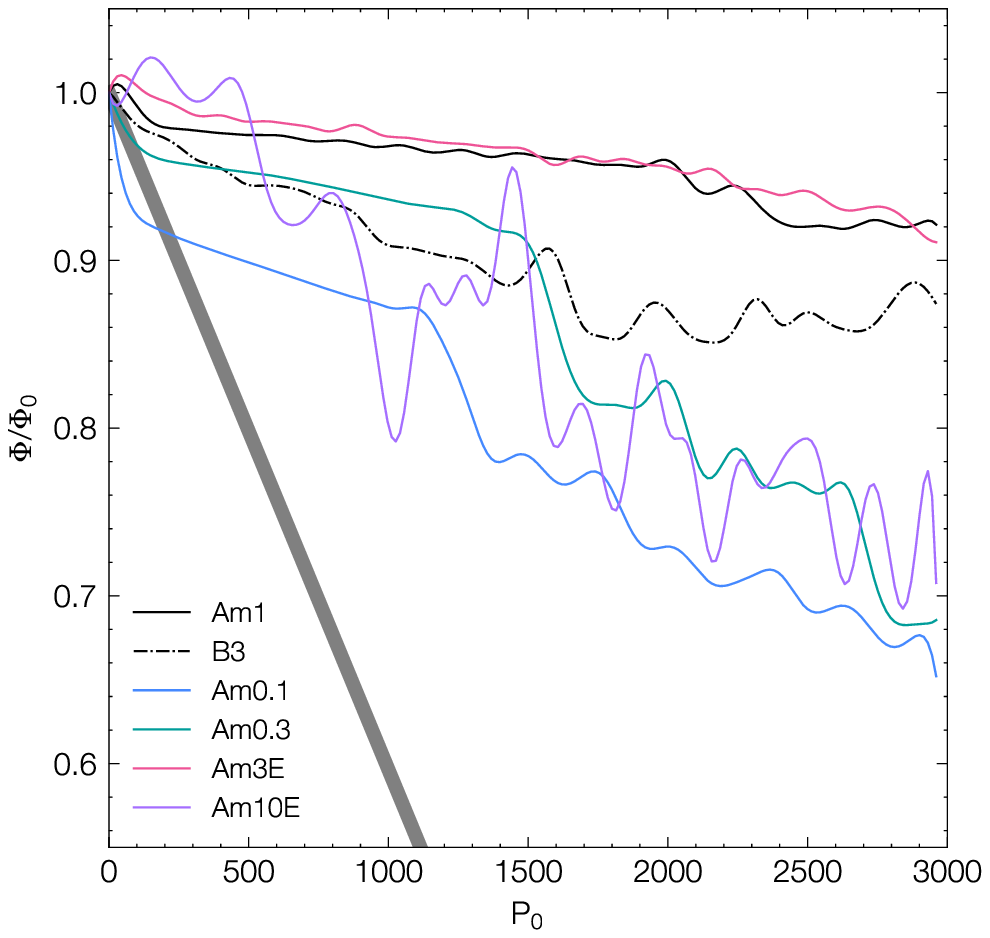}
\caption{The evolution of magnetic flux normalized by its initial value $\Phi/\Phi_0$ for all simulation models at $R=10$.}
\label{fig:FLUX}
\end{figure}

In the second panel of Figure \ref{fig:ct}, we see that within the disk, the vertical magnetic flux accumulates, assembling into magnetic flux sheets at separated radii. This phenomenon, known as magnetic flux concentration, has been reported in a variety of MHD simulations for PPDs (e.g. \citealp{bs14,bai15,bethune_etal17,suriano_etal18,suriano_etal19,rl19,rl20}). We note that, however, these existing studies are either local simulations, or 2D and 3D global simulations without properly resolving the MRI. Thus, our simulations can serve to provide robust evidence of the magnetic flux concentration phenomena, which will be analyzed in further details in this subsection and \S\ref{sec:Bpara}. 

In Figure \ref{fig:st}, we show the evolution of $B_z/B_{z0}$ at the disk midplane in a space-time diagram, where this quantity is azimuthally and vertically ($\pm1 H$) averaged.\footnote{There is largely no difference if we simply show $B_z/B_{z0}$ at the midplane without vertical averaging. With averaging, it better illustrates that flux concentration spans over a wide vertical range.} The overlaid white contours delineate linearly equally spaced lines of constant magnetic flux $\Phi(r,\pi/2)$ at the midplane.
From Figure \ref{fig:ct} and Figure \ref{fig:st}, we see that the development of magnetic flux sheets is associated with regions where the MRI turbulence is developed. As such regions expand towards larger radii over time, so do regions of magnetic flux concentration.
At $3000P_0$, discernible flux sheets emerge at $R \simeq 5.5, 8.5, 11, 13$, with high level of flux concentration.
In the bulk disk, the peak vertical field strength reaches a factor of $\sim4$ compared to its initial value, whereas 
in between the flux sheets it can be reduced to close to zero. This is also manifested in the radial profile of plasma $\beta$ in the top left panel of Figure \ref{fig:fig8}, where local minimum of $\beta$ are obtained at the locations of flux concentration.
We also see from the space-time diagram that the flux sheets are dynamically evolving in our fiducial Model \textsc{Am1}, where thicker sheets can split into multiple thinner sheets, and thin sheets can also merge.  

In our fiducial model \textsc{Am1}, the typical thickness of the flux sheets is $\sim0.5H$, and the flux sheets are typically separated by $\sim 2-5H$. 
Local shearing-box simulations that resolve the MRI turbulence show comparable thickness of the flux sheets \citep{bs14,bai15,rl18}. However, the separation between flux sheets is more ambiguous in these shearing-box simulations due to their local nature, where flux sheets are either too closely spaced with a separation of $\sim0.5H$ in unstratified simulations \citep{bs14}, or form only one single flux sheet over a simulation domain of several $H$ \citep{bai15}. Such ambiguity is resolved in our global simulations.

In Figure \ref{fig:Sig}, we further show the radial surface density profile $\Sigma(R)$ at $t=0,600,1200,1800,2400,3000P_0$, as well as radial profiles of $B_z/B_{z0}$ at the midplane for all models.
Focusing on the fiducial model \textsc{Am1}, we see that the overall surface density profile deviates from the initial profile, and develops multiple density maxima and minima. In particular, the surface density minima are co-located with where magnetic flux is concentrated, in agreement with previous local and global simulations (e.g. \citealp{bs14,bai15,bethune_etal17,suriano_etal19,rl20}). The width of the rings or gaps is $w_\Sigma\sim1.5-2.5H$. The surface density variations, estimated based on Equation \eqref{eq:sdv}, span over $\delta_\Sigma\sim 15-50\%$ for Model \textsc{Am1}. We defer further discussion on the physical mechanisms that produce density contrast and magnetic flux concentration phenomena in \S\ref{ssec:sf} and \S\ref{sec:mech}.

We note that the inner disk is gradually depleted over time (see \S\ref{sec:am}), which in reality may affect the ionization and thermal structure of the disk. Nevertheless, we neglect such further complications in this study, and our simulations are terminated before depletion becomes drastic.

\subsubsection{Toroidal magnetic field evolution}\label{sec:dy}

We see from the middle panels of Figure \ref{fig:ct} that the poloidal fields in the disk region, especially the flux sheets, are not purely vertical, but become locally tilted in a random manner. As the sign of toroidal field anti-correlates with the sign of the radial field component, the mean toroidal fields show arbitrary sign flips across the disk. The orientation of these tilts are dynamically evolving together with the flux sheets. This is accompanied by sign changes in toroidal field.
In the second column of Figure \ref{fig:fig8}, we see that the toroidal magnetic fields dominate over the poloidal components across the disk, with sign flips of $B_\phi$ also identified as a function of radius at the disk midplane.

In Figure \ref{fig:DY}, we show the time evolution of the vertical profiles of azimuthaly-averaged toroidal magnetic fields $B_\phi$ at $r=5.5$. The space-time diagram is often employed in the context of MRI dynamo actions, where the mean toroidal fields flip every $10$ orbital periods in ideal MHD simulations of the MRI with zero net vertical magnetic flux (e.g., \citealp{davis+10,shi+10,flock+11}), but the dynamo actions are weakened or even suppressed by net vertical magnetic flux with or without the presence of non-ideal MHD effects \citep{bs13a,simon_etal13a,simon_etal13b}.
The pattern in the space-time diagram of $B_\phi$ seen in Model \textsc{Am1} is clearly not a standard MRI dynamo, but rather just reflects the orientation of the tilt in poloidal fields. The sign flip at about $1850P_0$ is mainly associated with a change from $B_R>0$ to $B_R<0$ in the bottom half of the disk. Comparing to local simulations with similar setup \citep{simon_etal13a}, our results share similar irregular sign changes in the bulk disk, but lacks dynamo-like sign flips in the FUV layer ($\theta>\theta_{\rm trans}$). This is likely due to global conditions imposed by wind-launching from the disk surface, fixing the tilt of poloidal field and hence sign of $B_\phi$.

\subsubsection{Magnetic Flux Transport}\label{sec:ft}

Besides magnetic flux concentration, we further examine global trends of flux evolution. This is known as magnetic flux transport. As disk evolution is largely controlled by the amount of magnetic flux threading the disk, flux transport governs the long-term evolution of the disk at a more fundamental level. The white contour lines in Figure \ref{fig:st} show the time evolution of constant magnetic flux surface. It can be seen that after a rapid initial relaxation, the magnetic flux is slowly transported radially outward at a steady rate. Later on, as flux concentration develops, the location of constant flux surface tracks the location of the flux sheets where most of the flux assembles.

To quantify the radial transport of poloidal magnetic flux in Model \textsc{Am1}, we show the evolution of $\Phi/\Phi_0$, where $\Phi_0=\Phi|_{t=0}$, for all simulation models at $R=10$ in Figure \ref{fig:FLUX}. It illustrates the fraction of magnetic flux enclosed within $R=10$ compared to initial flux distribution. As we have discussed, the MRI is fully developed at this radius after $t=2000P_0$. Before this time, the enclosed magnetic flux steadily declines, corresponding to outward flux transport. Assuming the flux transport outward with such a steady rate, it requires $\sim800$ local orbits to lose half of the initial magneitc flux enclosed within $R=10$ for the fiducial model.
We have verified by performing additional axisymmetric 2D simulations with identical simulation setup 
and found the same rate of flux transport. This rate is much lower than that reported in \citet{bs17} (only about $\sim10\%$), which we speculate is related to the different simulation setup. In particular, wind in \citet{bs17} is launched from a lower height with $\theta_{\rm trans}=0.3$,  and it is found that the rate of flux loss is sensitive to $\theta_{\rm trans}$ under the current setup, where small transition angles correspond to fast magnetic flux losses (Yang \& Bai, to be submitted).

After $t=2000P_0$, magnetic flux transport is affected by flux concentration. For our fiducial model \textsc{Am1}, we see that the rate of flux transport stalls after $t\sim2400P_0$. This is associated with the fact that despite exhibiting dynamical evolution at earlier time, the locations of major flux sheets appear to be relatively steady. As a large fraction of magnetic flux is trapped in flux sheets, they serve to protect the system from further losing magnetic flux. On the other hand, we find a diversity of behavior from other models, to be discussed in Sections \ref{sec:Bpara} and \ref{sec:ind}.

\subsection{Turbulence and Stress Levels}\label{sec:tu} 

To parameterize  turbulence levels induced by the MRI, we compute the root mean square of velocity fluctuations, 
\begin{equation}
\delta v =\Bigg[\frac{1}{n}\sum\limits_{i=1}^n(v_i-\langle v \rangle)^2\Bigg]^{1/2},
\end{equation}
where angle bracket $\langle v \rangle$ denotes azimuthal averaging. The third column, top panel of Figure \ref{fig:fig8} presents the mean $\delta v/c_s$ as a function of radius, where the spatial average of $\delta v/c_s$ is performed over $\Delta R \in [0.95R, 1.05R]$ and $\Delta z \in [-3H, 3H]$, and the time average is over the last $400P_0$.
The turbulence is dominated by radial and azimuthal components, with  strengths of $\delta v\sim 0.02c_s $ for the radial extent $5< R <10$. The turbulence levels gradually decline over $R > 13$ as the MRI has not yet fully saturated in this region. 
The middle and right panels show the vertical profile of $\delta v$ at $R=5.5$ (with flux concentration) and $R=6.5$ (with less flux concentration), averaged
over $\Delta R \in [0.95R, 1.05R]$ and the last $400P_0$.
The turbulence strength is on the order of $0.01c_s$ towards the disk midplane and rise towards high latitude approaching $0.1c_s$ at the surface layers at both radii. 
This may suggest a turbulent FUV layer as seen in local shearing-box simulations \citep{simon_etal13b}.
The difference between regions with and without flux concentrations is moderate, which can be up to a factor of $\sim 2$ in $\delta v/c_s$.

The forth column of Figure \ref{fig:fig8} show the Reynolds and Maxwell stresses defined right below Equation \eqref{eq:alp}. The quantities are temporally and spatially averaged in the same way as for turbulence levels $\delta v/c_s$. 
We see that Maxwell stress dominates the Reynolds stress over the entire radial extent. The dimensionless $\alpha$ parameter is a few times $10^{-4}$ for Reynolds stress and $\sim 1-2 \times 10^{-3}$ for Maxwell stress in the radial domain $5< R <10$. 
Similar to turbulence levels, the difference between regions with and without flux concentrations of stresses is also moderate, up to a factor of $\sim 2-3$ in $\alpha$ parameter.

\subsection{Angular Momentum Transport}\label{sec:am}
 
\begin{figure}
\centering
\includegraphics[width=0.5\textwidth]{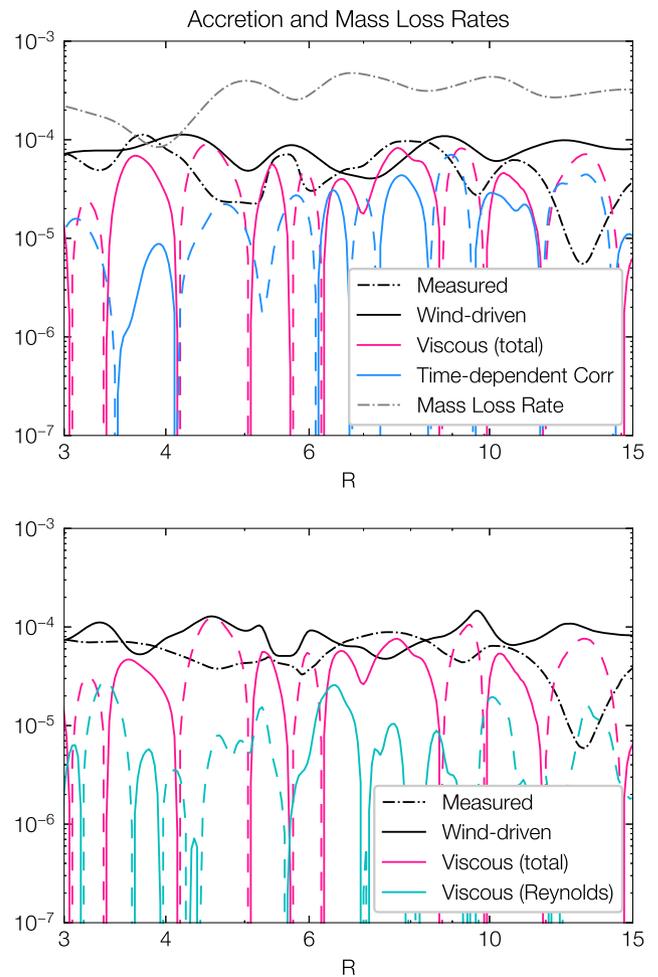}
\caption{Radial profiles of mass accretion and mass loss from the fiducial simulation, averaged from the last 300$P_0$. Top panel: contribution from individual terms based on the original decomposition in Equation (\ref{eq:am1}), as well as the mass loss rate. Bottom panels: actual accretion rate $\dot{M}_{\rm acc}$ by multiplying a correction factor ${\rm Corr}(R)$ to the top panel. See \S\ref{sec:diag} for definitions. Dashed curves denote negative accretion rates.} 
\label{fig:AM}
\end{figure}

\begin{figure}
\centering
\includegraphics[width=0.5\textwidth]{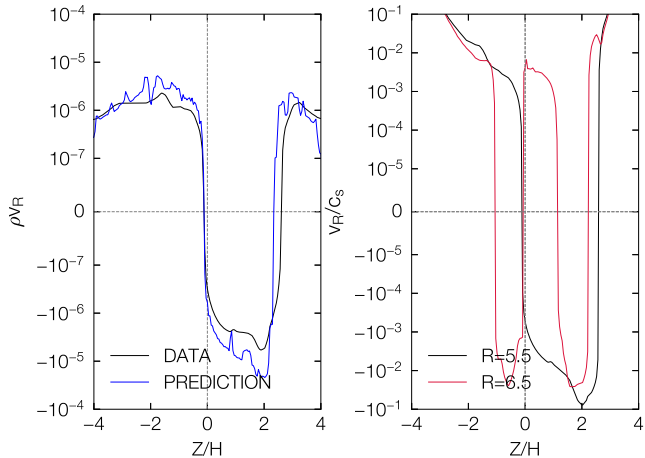}
\caption{Left: vertical files of mass flux at $R=5.5$ of Model \textsc{Am1}. Black curve denotes mass flux obtained directly from simulation data. Blue dashed curve denotes predictions by wind-driven accretion theory using Equation \eqref{eq:mf}. Right: vertical profile of radial velocity normalized by local sound speed at $R=5.5$ (black) and $R=6.5$ (dashed blue). The quantities are averaged over last $200P_0$ for both panels.}
\label{fig:mf}
\end{figure}

\begin{figure*}
\includegraphics[width=1\textwidth]{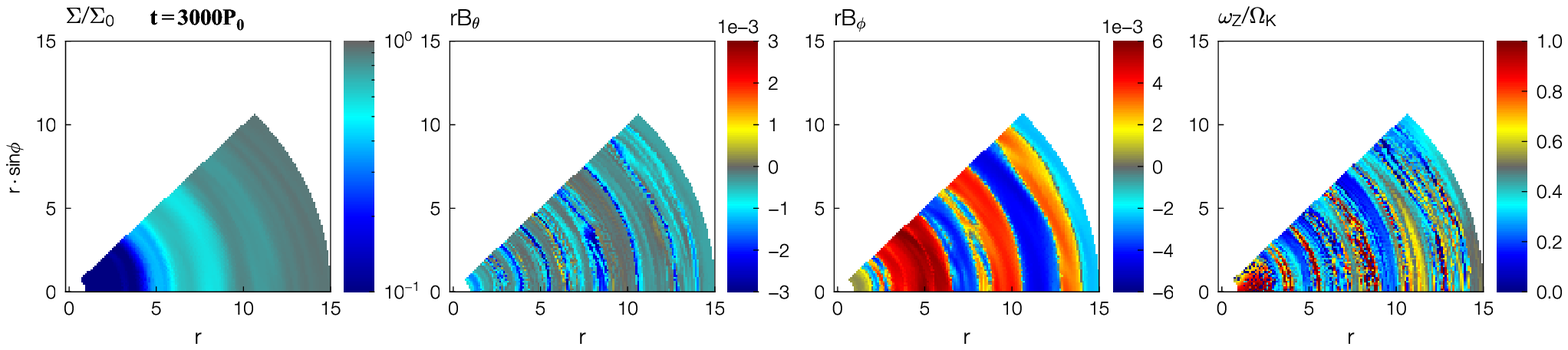}
\caption{Snapshots of surface density, meridional, azimuthal magnetic field, and vorticity at the midplane in ($R-\phi$) plane for Model $\textsc{Am1}$ at $t=3000P_0$. First panel: snapshots of normalized surface density normalized by its initial value $\Sigma/\Sigma_0$. Second panel: snapshots of scaled meridional magnetic field $r^2 B_\theta$. Third panel: snapshots of scaled toroidal magnetic field $r^2B_\phi$. Last panel: snapshots of vorticity normalized by Keplerian angular velocity $\omega_z/\OmK$. }
\label{fig:Rphi}
\end{figure*}

We discuss angular momentum transport and mass loss in the fiducial model in Figure \ref{fig:AM}. Following Section \ref{sec:diag}, we show in the top panel the ``original" decomposition of terms in Equation \eqref{eq:am1}, which reflects the torques exerted by various transport mechanisms. The data are averaged over the last 300$P_0$, except for the time-dependent term, which is computed from the finite difference between data from the first and second halves over this period, and gives an approximate estimate on the sign and magnitude of the term over a timescale of $\sim150/R^{3/2}$ local orbits. At each radii, we have further smoothed the data over $\pm0.5H$.

The accretion rate is on the order of $10^{-4}$ in code units, and is generally consistent with that expected for wind-driven accretion. Recalling Equation (\ref{eq:mdot_conv}), this accretion rate corresponds to $\gtrsim10^{-8}M_{\odot}$ yr$^{-1}$ in physical units for typical outer disks. We see that the dominant contribution is from vertical transport by the wind, but radial transport, as well as the time-dependent term, also play a significant role, and can reach a level comparable to wind-driven accretion at certain locations. Moreover, Maxwell stress is found to be a major contribution to the viscously-driven accretion, and the Reynolds stress has typically negligible contribution at all locations, consistent with expectations discussed in Section \ref{sec:tu}.

There is very significant mass loss in our simulations, with $d\dot{M}_{\rm loss}/d\log R$ comparable or even exceeding the accretion rate. Such significant mass loss is generally expected for magneto-thermal disk winds \citep{bai+16}, where thermal pressure plays an important role and has been observed in a number of simulations of the kind (e.g., \citealp{bs17,cb20,rodenkirch20}). On the other hand, the exact mass loss rate is sensitive to the thermal chemistry in the wind launching region, where our treatment is considerably simplified, and more realistic simulations incorporating these physics tend to yield more mild mass loss rates \citep{wang_etal19,gressel_etal20}. Therefore, when discussing wind mass loss, we focus on the trends instead of absolute values.
The significant mass loss can alter the surface density profile, potentially leading to the depletion of the inner disk (e.g., \citealp{suzuki_etal16}), and this is indeed seen in our simulations. 
Nevertheless, we will not pursue further discussion on this density depletion due to unrealistic mass loss in our simulations.

\subsubsection{Substructure formation}\label{ssec:sf}

Magnetic flux concentration leads to variations in wind-driven accretion and mass losses. We see that the wind stress $B_zB_\phi$ is directly correlated with magnetic flux distribution when comparing with the second panel of Figure \ref{fig:Sig}, where regions with flux concentration has on average stronger stress. However, the radial variation in the $B_zB_\phi$ stress is not as dramatic as magnetic flux concentration seen in the midplane. This is readily visible from the middle row of Figure \ref{fig:ct}, where magnetic flux becomes less concentrated in the surface layer.
The distinction of magnetic flux concentration in the AD-dominated regime from the ideal MHD regime was observed in local shearing-box simulations \citep{bs14}, where only the former leads to thin flux sheets (see also our discussion in Section \ref{sec:Bpara}). Therefore, we expect the thickening of flux sheets in the wind zone is likely associated with the transition into the ideal MHD regime.

Interestingly, we notice that peaks in the measured mass loss rate profile do not coincide, but tend to anti-correlate with those in the stress profile. This is somewhat counter-intuitive as one generally expects stronger net vertical field to yield both higher wind-driven accretion and mass loss rates. This expectation is based on the assumption that gas is well coupled to magnetic field and hence its streamlines follows poloidal magnetic fields. As we implement the transition from non-ideal MHD to ideal MHD regime over a relatively broad ($\sim H$) layer, we do observe departures between flow streamlines and poloidal field lines over the transition region. More systematic studies are needed to examine how the mass loss profile depends on the smoothness of this transition.

The formation of ring-like substructures further complicates the process of angular momentum transport. They are more strongly exhibited in radial transport of angular momentum, as well as the time-dependent correction term. Because radial transport of angular momentum depends on the radial gradient of the $R\phi$ stress, radial variations in surface density as well as the $\alpha$ parameter seen in Figure \ref{fig:fig8} give rise to frequent sign changes in its contribution to gas accretion. In addition, the mass accumulation and/or dilution in rings/gaps also changes the level of radial pressure gradient and hence rotation profiles, which is at the expense of angular momentum and leads to the first term in Equation (\ref{eq:am1}), i.e., the time-dependent correction. Contribution from these terms, coupled with ring-like substructure, give rise to additional variations in the total accretion rates, as can be readily tracked in Figure \ref{fig:AM}. These terms could be considered as non-linear back-reaction to disk angular momentum transport in response to driving from magnetic-flux concentration.

The bottom panel of Figure \ref{fig:AM} shows the actual accretion rate profiles, as well as contributions from individual terms. This is obtained by correcting for the $\partial j/\partial R$ factor. We see that after this correction, there are more frequent radial variations at relatively small amplitudes in wind-driven accretion rates, but they barely show any correlation with the profile of poloidal magnetic flux. In fact, the amplitudes of rings and gaps are dynamically evolving, reflecting the complex interplay among different transport mechanisms together with the aforementioned ``back-reaction" due to formation of disk substructure. Altogether, we see that despite substantial magnetic flux concentration in the bulk disk, the disk rarely form deep gaps.

\subsubsection{Vertical profiles of the accretion flow}

In the left panel of Figure \ref{fig:mf}, we examine the vertical structure of the accretion flow by showing the azimuthally and temporally averaged radial mass flux profiles at $R=5.5$ for the fiducial Model \textsc{Am1}. The quantities are averaged over last $200P_0$.
We see there are both accretion and decretion flows of comparable level of mass flux present at different heights, thus the net accretion flow that we discussed earlier results from cancellations from such flows. By comparing the results with Equation (\ref{eq:mf}), we see that the mass fluxes are generally consistent with predictions
suggesting that the overall flow structure is primarily driven by the torques exerted by magnetic forces. In particular, it is set by the vertical gradient of the toroidal field, and following the discussion in Section \ref{sec:dy},
the locations of the accretion/decretion flows are random and dynamically-evolving. 

We also see that the mass flux profiles show departures from predicted by Equation (\ref{eq:mf}). Such departures can be substantial, sometimes a factor of a few in the peaks. This reflects the role played by the MRI turbulence. Following our previous discussion, contributions to such departures include radial transport and time-dependent correction, which is in part associated with formation of disk substructures.

The right panel of Figure \ref{fig:mf} illustrates the radial velocity normalized by local sound speed as a function of height, at radii of $R=5.5$ and $R=6.5$,
showing alternating radial velocity on the order of $\sim 10^{-2}-10^{-1}$ within $\pm 2H$.
We anticipate that such alternating radial gas flows driven by disk winds can effectively enhance the dust radial diffusion due to the weak MRI turbulence alone, as shown in \citet{hb21}. It would be interesting to explore this subject further by incorporating a dust component, as the level of radial diffusion in dust, together with the profile of the gas rings, sets the width of dust rings in mm-continuum observations \citep{dullemond+18}.

\subsection{Azimuthal Features}\label{sec:af}

The discussions above are based upon azimuthally averaged simulation data. It remains to address whether these features are largely axisymmetric or involve additional non-axisymmetric structures.

In Figure \ref{fig:Rphi}, we show snapshots in ($R-\phi$) plane of surface density, scaled meridional and azimuthal magnetic fields, and vorticity at the midplane. 
In the second panel, the meridional field $B_\theta$ is mainly axisymmetric, indicating that the vertical magnetic flux is indeed confined in flux sheets rather than an artifact of azimuthal averaging. 
Likewise, the toroidal field $B_\phi$ shown in the third panel is also highly axisymmetric, which exhibits sign changes in $R$, similar to sign changes in $z$ and time as previously discussed. We note that toroidal fields reverse polarity sharply, leading to steep gradients at certain radii. This phenomenon could be attributed to the sharp structures resulted from the anisotropic diffusive properties of AD \citep{bz94}.

The formation of ring-like substructures may induce vortex formation by Rossby wave instability (RWI) \citep{lovelace99,li_etal00}. In the last panel, we exmamine the $z-$component of midplane vorticity, defined as
\begin{equation}
\omega_z = (\nabla \times \bfv)_z, 
\end{equation}
where axisymmetry and Keplerian rotation give $\omega_z/\OmK=0.5$. We note that no discernible vortex formation is observed, indicating the absence of the RWI. \citet{ono_etal16} calculated the fastest growth rates as a function of amplitudes and widths of the surface density variation (see their Figure 6). Taking the Gaussian bump model for surface density variations in their paper, and given a width of $w_\Sigma\sim1.5-2.5H$ for our fiducial Model \textsc{Am1},
the smallest amplitude of surface density variation to trigger the RWI is on the order of $\sim100-200\%$, which is achieved with the azimuthal wavenumber $m=1$. Seen in our Figure \ref{fig:Sig}, fiducial model \textsc{Am1} results in variations of $\delta_\Sigma\sim15-50\%$. Despite our simulations have a reduced azimuthal domain, the critical surface density variations found in \citet{ono_etal16} is even higher for larger azimuthal wavenumber $m$. This is consistent with the lack of vortices in the last panel.

\section{Parameter Study}\label{sec:pm}

\begin{figure*}
\centering
\includegraphics[width=0.95\textwidth]{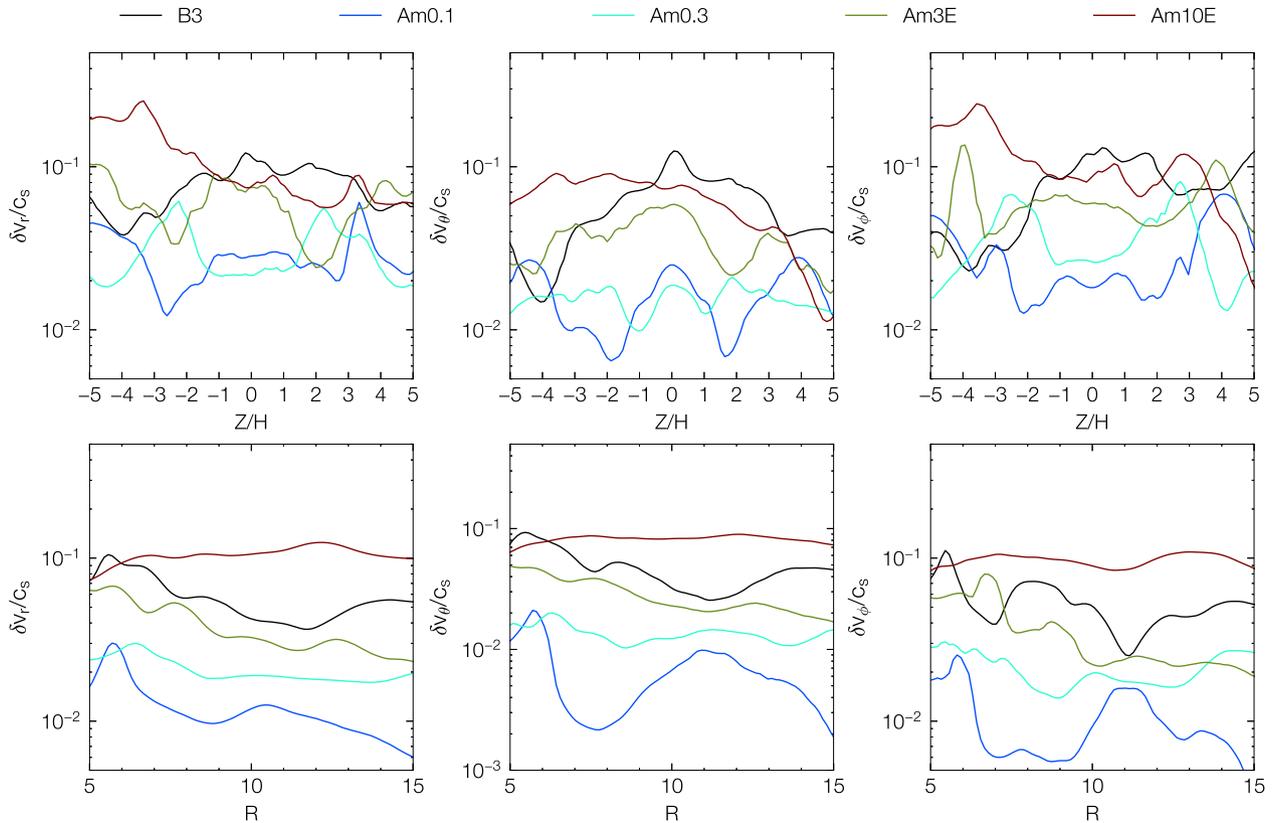}
\caption{Turbulence levels $\delta v_r, \delta v_\theta, \delta v_\phi$, normalized by local sound speed $c_s$, as a function of height $z$ at r=5.5 (top row) and of cylindrical radius $R$ (bottom row) for Models \textsc{B3, Am0.1, Am0.3, Am3E, Am10E}.}
\label{fig:turbpara}
\end{figure*}

\begin{figure}
\centering
\includegraphics[width=0.5\textwidth]{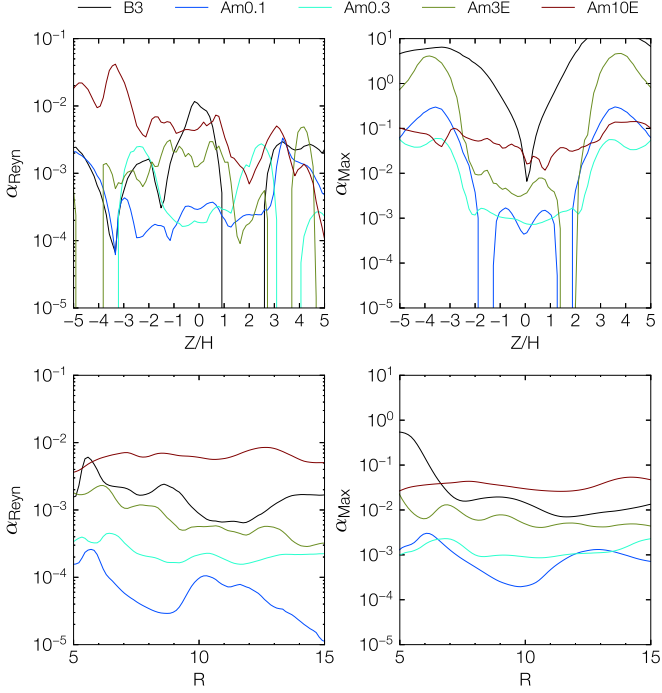}
\caption{Reynolds and Maxwell stresses $\alpha_{Reyn},\alpha_{Max}$ as a function of height $z$ at r=5.5 (top row) and of cylindrical radius $R$ (bottom row) for Models \textsc{B3, Am0.1, Am0.3, Am3E, Am10E}.}
\label{fig:turbpara_r}
\end{figure}

\begin{figure}
\centering
\includegraphics[width=0.45\textwidth]{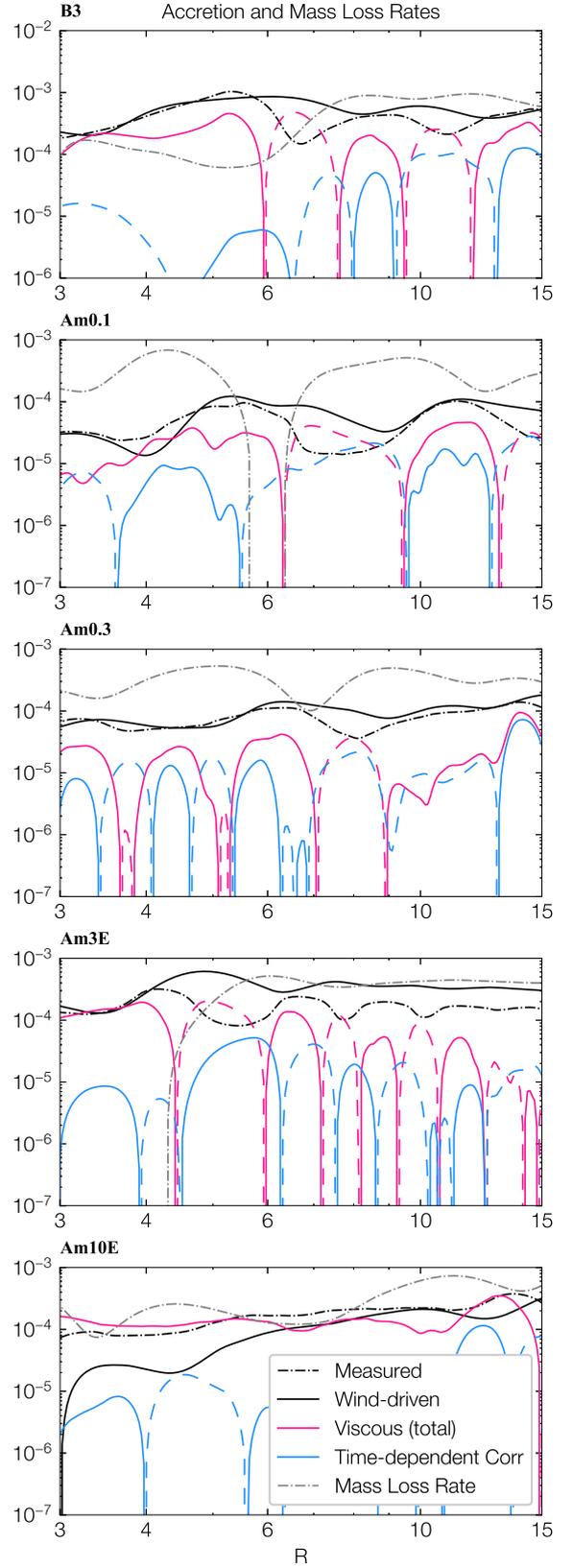}
\caption{Radial profiles of mass accretion and mass loss from all other simulations (excluding the fiducial run). Shown are the contribution from individual terms based on the original decomposition in Equation (\ref{eq:am1}), as well as the mass loss rate. They are the counterpart of the top panel of Figure \ref{fig:AM}. Dashed curves denote negative accretion rates.} 
\label{fig:AMpara}
\end{figure}

\begin{figure}
\includegraphics[width=0.5\textwidth]{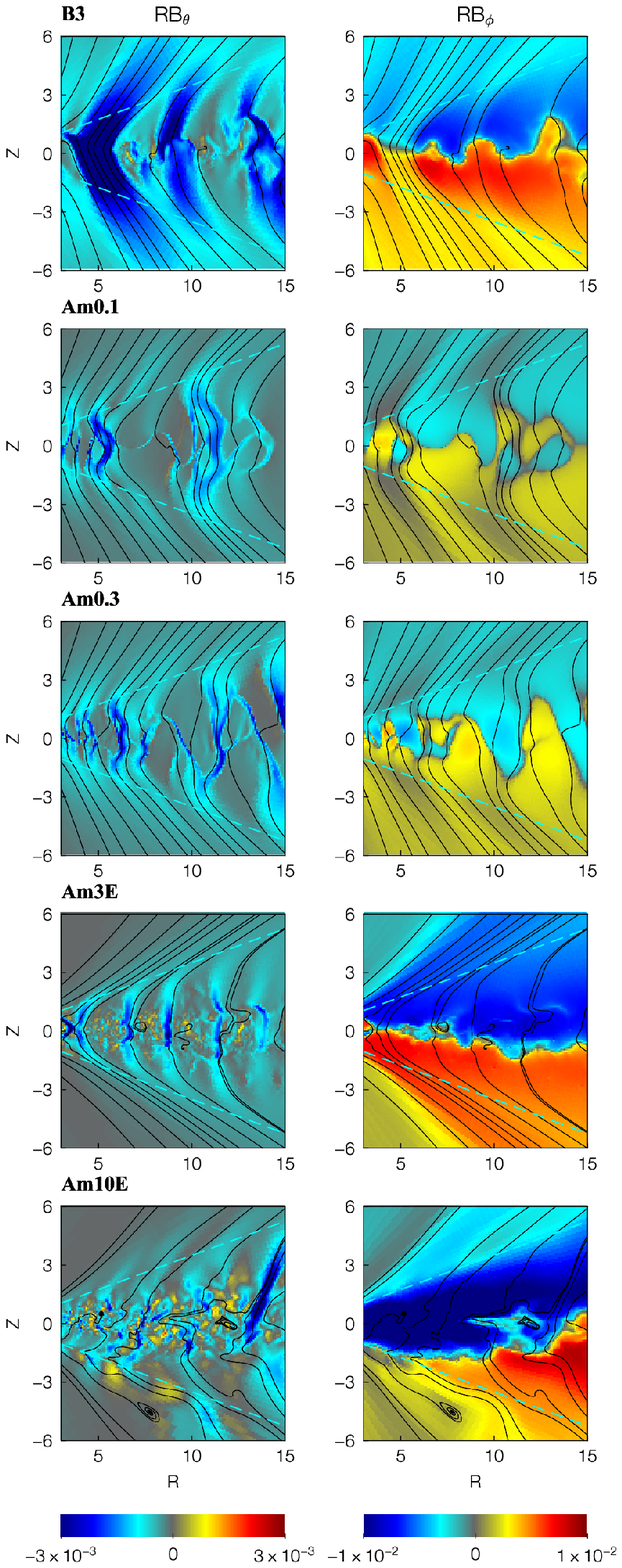}
\caption{Left: snapshots of scaled meridional magnetic field $RB_\theta$ at $t=3000P_0$. Right: snapshots of scaled toroidal magnetic field $RB_\phi$ at $t=3000P_0$. Overlaid black curves are equally spaced contours of poloidal magnetic flux.}
\label{fig:Bpara}
\end{figure}

In this section, we present the results for Model \textsc{B3, Am0.1, Am0.3, Am3E, Am10E},  where \S\ref{sec:flpara} focuses on flow properties including turbulence levels, stresses, and angular momentum transport, \S\ref{sec:Bpara} discusses magnetic flux concentration and transport, and \S\ref{sec:ind} details each individual simulation model.

\subsection{Flow properties}\label{sec:flpara}

Similar to Figure \ref{fig:fig8} for the fiducial model \textsc{Am1}, we show in Figure \ref{fig:turbpara} and Figure \ref{fig:turbpara_r} the vertical and radial profiles of turbulence levels and stresses for the rest of the simulation models. The averaging is taken in the same way as in Figure \ref{fig:fig8} (see \S\ref{sec:tu}). In Figure \ref{fig:turbpara}, we see that the turbulence levels increase with larger $Am$ and stronger magnetization (smaller $\beta_0$). The values of $\delta v/c_s$ gradually raise from $\sim 10^{-2}$ for Model \textsc{Am0.1} to $\sim 10^{-1}$ for Model \textsc{10E} as well as for Model \textsc{B3}. Correspondingly, the Reynolds stresses $\alpha_{R\phi}$ increase from $\sim 10^{-4}$ to $\sim10^{-3}$, and the Maxwell stresses from $\sim 10^{-3}$ to $\sim10^{-1}$. Unlike the fiducial model \textsc{Am1}, we do not observe significant boost of turbulence levels at the wind region for the rest of simulation models, but rather, turbulence level is largely flat over height.

Observations on turbulent motions through non-thermal line broadening infer a turbulent velocity $\delta v<0.04-0.06c_s$ at different heights \citep{flaherty_etal17}. The vertical profiles of $\delta v/c_s$ shown in Figure \ref{fig:fig8} and Figure \ref{fig:turbpara} hence suggest our models with $Am\leqslant 1$ for $\beta_0=10^4$ are favored, which place less stringent constraints than in \citet{simon_etal18}.

Figure \ref{fig:AMpara} shows radial profiles of accretion rates with contributions from Equation (\ref{eq:am1}), similar to the top panel of Figure \ref{fig:AM}. Angular momentum transport in these models is dominated by magnetized winds, consistent with the fiducial model. Accretion rates are largely set by the net vertical magnetic flux, and is similar among models with $\beta_0=10^4$. Model \textsc{B3} possesses an accretion rate with a factor of $\sim10$ higher as expected, since accretion rate scales quadratically with field strength. Contributions to accretion rates from individual terms also show radial variations akin to magnetic flux concentration. The mass loss rates remain substantial, with ratio of $d\log\dot{M}_{\rm loss}/d\log R$ to accretion rates reaching a few. This ratio is smaller in Model \textsc{B3}, as is expected in disk wind theory for stronger magnetization
\citep{bai+16}. Radial transport by Maxwell stress and time-dependent correction show frequent sign changes that are associated with disk substructures. In sum, these models share qualitatively similar features to those in the fiducial model (\S\ref{sec:am}).

\subsection{Magnetic flux evolution}\label{sec:Bpara}

All of these simulation models show features of spontaneous magnetic flux concentration (Figure \ref{fig:st} and Figure \ref{fig:Bpara}), though their properties, such as the amount of flux sheets emerged, rates of outward flux transport, widths of rings, and surface density variations, are diverse. All models show the formation of rings and gaps in surface density profiles due to the flux concentration (Figure \ref{fig:Sig}). The widths of rings or gaps span over $w_\Sigma\sim 1-5H$, and the surface density variations cover $\delta_\Sigma\sim 30\%-200\%$ for different simulation models.

The space-time diagrams of toroidal magnetic field $B_\phi$ at $r=5.5$ are shown in Figure \ref{fig:DY}. Similar to the fiducial model \textsc{Am1}, the sign in $B_\phi$ mainly reflects the orientation of poloidal field through the disk. We note that the toroidal field shows symmetry about the midplane, only flipping sign once across the disk for Model \textsc{B3}, \textsc{Am3E}, and \textsc{Am10E}, while Model \textsc{Am0.1} and \textsc{Am0.3} show more stochastic sign changes. This trend could be related to the fact that for stronger net poloidal magnetic fields or weaker non-ideal MHD effects, $B_\phi$ is more strongly amplified, 
which is more stiff against reversal.

The overall transport of magnetic flux, as examined in Figure \ref{fig:FLUX}, show a two-stage evolution (except for Model \textsc{Am10E} which quickly turns into strong turbulence), with an initial steady decline before the development of the MRI and magnetic flux concentration. Interestingly, after the development of the MRI turbulence, for Model \textsc{Am1} and Model \textsc{B3}, the flux sheets become stationary and cease to drift outwards. For Model \textsc{Am0.1} and Model \textsc{Am3E}, magnetic flux keeps drifting outwards, but at an averaged rate comparable to the first stage.
Model \textsc{Am0.3}, on the other hand, exhibits faster outward transport with flux concentration. The lack of trend in our results likely reflects the complexity in magnetic flux transport in MRI-turbulent disks.

\subsection{Notes on individual simulations}\label{sec:ind}

In this subsection, we discuss special features in each individual simulation, owing to the diversity of our simulation results.

\textbf{Model B3} -- with $\beta_0=10^3$, this model is most similar to several previous global disk simulations that report magnetic flux concentration \citep{bethune_etal17,suriano_etal18,rl20}. 
Comparing to previous global simulations (e.g. \citealp{suriano_etal19}), the flux sheets in this simulation are more incoherent and short-lived since the MRI turbulence is resolved. We speculate that it is related to the corrugation of midplane current sheet seen in Figure \ref{fig:DY} that oscillates over time. 
The flux sheets are also thicker than those in the fiducial model \textsc{Am1}, which may be due to stronger turbulent diffusion that likely act against flux concentration.

Three prominent thick flux sheets are present at the end of simulation as seen in Figure \ref{fig:Bpara}, resulting in the formation of three gaps in Figure \ref{fig:Sig}. The gaps located at $R\gtrsim 6$ have widths $w_\Sigma\sim3H$ in surface density, and a surface density variable of $\delta_\Sigma\sim20\%-70\%$. We note that a thick flux sheet is developed at $R\sim 4-6$, driving a wide deep gap. As this region is close to the inner boundary, the gas surface density is significantly depleted over simulation timescale, attributed to the fast accretion and significant mass loss. 
Hence, the disk is more heavily magnetized.
This depletion with trapped magnetic flux merits more in-depth study in the future.

\textbf{Model Am0.1} --
With $Am=0.1$, this model is borderline since the MRI transitions to the ambipolar shear instability in the limit of $Am\ll1$ \citep{kunz08,pw12}. 
In Figure \ref{fig:Bpara}, we see that two major concentrated flux sheets are developed, locating at $5<R<6$ and $10<R<15$. These two sheets are almost stationary, promoting efficient mass depletion in these regions, leading to relatively deep gaps. Seen in Figure \ref{fig:Sig}, the surface density shows two bumps with widths of $w_\Sigma\sim4-5H$ and varies by $\delta_\Sigma\sim80\%-200\%$. We have also experimented with different inner boundary conditions (e.g., whether to enforce $E_\phi=0$) while keeping $Am=0.1$, finding that the small $Am$ simulations tend to generate fewer but widely-separated flux sheets regardless of the numerics.
We also note that in Figure \ref{fig:AMpara}, regions with strong magnetic flux concentration show reduced or negative mass loss rate. This is related to the fact that the fiducial location where we set as the wind base is still occupied by the accretion flow (with negative vertical velocity) in this model.

\textbf{Model Am0.3} -- 
The flux concentration phenomenon present in this model shares similarities with Model \textsc{Am1} and Model \textsc{Am3E} (Figure \ref{fig:Bpara}), where multiple concentrated flux sheets are emerged. Two discernible bumps seen in surface density (Figure \ref{fig:Sig}) are associated with two sheets with stronger flux concentration located at $R=6$ and $R=11$. The corresponding widths of rings and gaps are $w_\Sigma\sim3-5H$, and the surface density varies by $\delta_\Sigma\sim80\%-130\%$. In Figure \ref{fig:st}, the flux sheets show fastest migration towards large radii among all simulation models, accompanied by rapid flux transport as seen in Figure \ref{fig:FLUX}.

\textbf{Model Am3E} -- 
Seen in Figure \ref{fig:st} and Figure \ref{fig:Bpara}, the flux sheets in this model are more ordered and particularly steady comparing to other models.
These thin flux sheets correspond to a narrower width of $w_\Sigma\sim1-2H$ in surface density, and a surface density variable of $\delta_\Sigma\sim30\%-60\%$.
The magnetic flux is slowly leaking outward comparing to the fiducial model (Figure \ref{fig:FLUX}).
 
\textbf{Model Am10E} -- 
Model \textsc{Am10E} is close to the ideal MHD regime and is characterized by strongest turbulence among all the other models. As a result, while magnetic flux concentration can be identified, it appears flocculent in the space-time diagram and does not lead to appreciable substructures (Figure \ref{fig:st}).
The outward transport of flux is fast. The associated surface density does not display prominent annular substructures (Figure \ref{fig:Sig}).
In this model, the features of MRI dynamo are still not observed, in contrast with local shearing-box simulations for which $Am=10$ simulation results in a period of dynamo flipping of $B_\phi$ on the order of $50$ local orbits \citep{simon_etal13b}. 

\section{Discussion}\label{sec:ds}

\subsection{Numerics}\label{ssec:num}

\textbf{Grid and resolution} -- 
We emphasize that conducting 3D simulations with sufficiently high resolution to resolve the MRI turbulence is essential to properly characterize the gas dynamics of outer PPDs. 
We have experimented with 2D simulations %are experimented utilizing 
with the same resolution, and the results indicate that the system is steady with no evidence of MRI for the long term evolution. 
Our results can also be compared with recent global 3D simulations (e.g. \citealp{suriano_etal19,hu_etal19}) that employed coarser resolutions ($\sim 9$ cells per $H$), where the flows were considered to be almost laminar without addressing the development of MRI turbulence. On the other hand, the condition to properly resolve the MRI turbulence in non-ideal MHD is yet to be established, especially given the wide range of $Am$ values explored here. If we adopt the quality factor employed in ideal MHD simulations \citep{hawley+11} (see their Equation 10), and replace the most unstable wavelength $\lambda_{\rm MRI}$ with the fitting formula of \citet{bs11} (see their Equation 22), we find the condition in the azimuthal dimension is well satisfied with $Q_y\gg10$, and that $Q_z\approx10$ for most runs, indicating turbulence is marginally resolved in the vertical dimension.

One caveat of this work arises from the limited $\phi-$domain employed. More extended $\phi-$domain would accommodate more low-$m$ MRI modes to emerge, while restricted $\phi-$domain may artificially amplify turbulence, based on ideal MHD simulations of the MRI turbulence \citep{flock12}. At present stage, it is computationally infeasible to scan a large parameter space of $Am$ as in this paper while employing an extended azimuthal domain. We leave extended $\phi-$domain global simulations for a future study.

\textbf{Inner boundary condition} --
We have taken special care in setting the inner radial boundary conditions. As we focus on the gas dynamics in the outer PPDs, regions inside the inner boundary corresponds to the inner disk, which is expected to be threaded by poloidal magnetic flux and launches its own disk winds. As this region is artificially truncated by our inner boundary, one expects there is poloidal magnetic flux anchored to this boundary, and would ideally design the boundary conditions that reflect such conditions: it should accommodate mass injection mimicking inner disk outflows at high latitudes, as well as the conventional ``outflows" which represents accretion flow in the disk region.
Our choice of a ``fixed-state" type boundary condition represents a compromise that allows artificial launching of outflows from the high-latitude region of the rotating inner boundary, despite of having some mass accumulation in the disk region of the inner boundary which becomes less realistic after long-term evolution. This is to be compared with the conventional outflow-type boundary condition, which tends to pull gas into the boundary from high altitude,
and further attracts and destabilizes poloidal field lines near the inner boundary, although they would have been part of a more stable wind launched from the putative disk regions inside the inner boundary. In ideal MHD simulations, it leads to coronal accretion (e.g., \citealp{zhu18}), and as field lines penetrating to the inner boundary can extend to large distances, they may substantially affect the overall simulation results.
On the other hand, we note that coronal accretion still develops when modeling the inner boundary to mimic a weakly magnetized protostar \citep{takasao_etal18}. Such differences highlights the importance of inner boundary conditions in global disk simulations, and merits more systematic studies in the near future.

\textbf{Thermodynamics and chemistry} --
We have opted for simple thermodynamics, prescribing the temperature, ambipolar diffusivity, and thermal relaxation timescales. The excessive mass loss via MHD winds observed in all models could be possibly due to these crude prescriptions, and more detailed treatment including ionization thermochemistry and radiative transfer \citep{bai17,wang_etal19,gressel_etal20} shall be considered to properly track the wind launching, acceleration, and propagation.

\subsection{Mechanisms of magnetic flux concentration}\label{sec:mech}

A handful of mechanisms have been proposed to explain the spontaneous magnetic flux concentration seen in global simulations. Previously, \citet{bs14} speculated that in the ideal MHD regime, the recurrent action of the MRI channel flows makes oppositely directed fields induced by the channel flows reconnect, producing magnetic loops and separating mass from magnetic flux. 
\citet{suriano_etal18} found that at the disk midplane, toroidal magnetic field reverses polarity, promoting pinching of poloidal field lines and fast accretion. The reconnection of the pinched field acts to separate mass from magnetic flux. Lastly, \citet{rl19} demonstrated a linear secular instability where initial radial density perturbations lead to transport of gas and magnetic fluxes towards the gap via viscous stress. The excess of poloidal magnetic flux in the gap extracts more mass so that initial density perturbations are reinforced. 
The presence of the MRI turbulence and diversity of flux concentration phenomena in our simulations complicate the quest for concentration mechanisms. In the following, we give a qualitative phenomenological discussion on the possible causes focusing on the last two scenarios.

Our Model \textsc{B3} shares similarity with \citet{suriano_etal18}. Specifically, by closely examining the simulation data, we find small magnetic loops presumably generated by the magnetic reconnection are emerged at the current layers in between the segregated poloidal field lines. For the rest of our simulation models which all possess $\beta_0=10^4$, however, we do not observe such magnetic loops at the midplane, especially because that most of these models lack persistent midplane current sheet.

The secular wind instability requires, 
firstly, turbulence to be built in the disk in order to yield viscous stress. Secondly, the magnetic field shall be well-coupled to the gas such that magnetic flux is brought towards gaps together with gas. Lastly, stronger magnetization should lead to an excess of ejected mass through disk winds. 
Our simulations do possess excessive mass loss, but it is far from straightforward to test whether magnetic flux is brought to the gap by the viscous flow towards filling in the gap.
One possible way to testify this mechanism is by increasing the transition heights of the disk $\theta_{\rm trans}$, for which mass loss rate by wind can be reduced, and one can inspect if the spontaneous ring formation is weakened. 

\subsection{Implications for annular substructures}\label{sec:6.3}

Our simulation results reveal the fact that the spontaneous ring formation through magnetic flux concentration is robust over a wide range of parameter space, including different disk magnetization $\beta_0$ and ambipolar diffusion strengths.
Our results also suggest that magnetically-induced substructures in the MRI turbulence show a diversity of properties.
We have seen that the process is stochastic, and the spacing of the substructures to some extent sensitively depends on simulation parameters. Besides the simulations shown in this paper, we have also experimented with other treatments for the buffer zone near the inner radial boundary, including the strengths of resistivity buffer and enforcing $E_\phi=0$. Given the same physical parameters, these simulations share qualitatively similar disk evolutionary patterns and properties of flux sheets, such as stresses and turbulence levels as well as separations and field strengths of the flux sheets. Nevertheless, they differ case by case in terms of exact time and locations in developing these concentrated flux sheets, and their migrations over time.

Another important finding from our results is that the substructures are typically shallow, with surface density contrast being order unity or less, and dynamically evolving. For instance, in our Model \textsc{B3}, \textsc{Am1} and \textsc{Am3E}, the locations of rings and gaps in the radial surface density profiles are time-dependent (Figure \ref{fig:Sig}).
Since observations on ring-like substructures are through dust, it is unclear whether dusts could respond timely. In other words, such shallow substructures may evolve faster, a few tens of local orbital times, than the dust trapping timescales, precluding the formation of deep rings and gaps. 

Observationally, by assuming steady state with dust concentration balanced by turbulent diffusion, one can constrain the level of turbulence $\alpha/St$ by separately estimating the width of dust and gas ring width, where $St$ is the dust Stokes number, or dimensionless stopping time characterizing the level of coupling with gas. From the DSHARP disk sample \citep{andrews+18}, it was found that $\alpha/St\sim0.1$, implying non-negligible level of disk turbulence depending on the not-well-constrained dust Stokes number \citep{dullemond+18,rosotti+20}, with upper limit of up to $\alpha\sim10^{-2}$. Level of turbulence in our simulations easily satisfies these constraints, and in the meantime, our simulation results call for caution in this approach because the steady-state assumption may not hold.

While rings and gaps are commonly interpreted as a result of interactions between disks and the embedded planets 
(e.g. \citealp{pinte_etal20}), by resolving key disk microphysics, our simulations reveal that the mechanism drives disk angular momentum transport and evolution is inherently non-smooth, and could result in additional, likely shallower disk substructures. 
Our magnetic mechanism might be a plausible explanation for substructures found in
young Class 0/I disks, for which the evolutionary stage of these systems might be too early for Jupiter-mass planets to form \citep{se18,nakatani_etal20,segura-cox_etal20,sheehan_etal20}.
For more matured, Class II disks, \citet{jennings+21} employed a super-resolution technique and discovered finer disk substructures in the DSHARP disk samples, including shallow annular substructures in the outer disks. Our results could be a possible interpretation for such shallow substructures.
 
\subsection{Implications for B-field observations}\label{sec:6.4}

Observing magnetic fields in PPDs is challengin. Recent attempts have been made through the Zeeman effect, obtaining only upper limits \citep{vlemmings_etal19,harrison_etal21}. These upper limits constrain the instantaneous line-of-site magnetic field strength averaged over the line-emitting region, typically in the outer disk surface layers. Models used to constrain Zeeman observations commonly make simplified assumptions about magnetic field geometry \citep{brauer_etal17,mazzei_etal20}, which in our simulations is dominated by the toroidal component.
However, the toroidal field randomly reverses polarity in space and time within the bulk disk, which could potentially lead to cancellations of the Zeeman signal and hence complicates the interpretation of observations. 
On the other hand, we note that for the widely-considered molecule of the Zeeman effect, CN mainly traces disk heights of $z/r\sim0.2-0.4$ \citep{cazzoletti_etal18,teague_etal20}.
These locations might be sufficiently high so that the sign of toroidal field does not flip, as in our ficucial model \text{Am1}, seen in the second column, top panel of Figure \ref{fig:fig8}. Nevertheless, future works with more realistic treatment of thermodynamics in the wind launching region shall be conducted in order to assess whether there can be sign flips of $B_\phi$ in the CN-emitting regions.

Another approach to probe disk magnetic field is through paleomagnetism, by 
recovering remnant field recorded in meteorites when they formed in the solar nebula
\citep{wbf21}, usually believed to be in the disk midplane region.
One major uncertainty in paleomagnetic measurements is that the radial locations where meteorites parent bodies formed are largely unknown, leaving a broad range of plausible radii in the solar nebula. Comparison to theories in the literature have relied on models that assume the disk magnetic field strength is smooth, and link field strength to expected disk accretion rates
(e.g. \citealp{wardle07,BG09,wbf21}). Fortunately, although poloidal magnetic fields become spontaneously concentrated into flux sheets, the total field strength dominated by toroidal field is relatively smooth, as seen in the second column of Figure \ref{fig:fig8}, which validates this approach.

By scaling numerical units in the simulations to physical units of magnetic fields using Equation (\ref{eq:B_conv}), we find a magnetic field strength of $\sim 0.01$ G for typical outer disk conditions around $\sim30$ AU from our simulation region between $R\sim5-10$. 
We can verify the basic principles used in paleomagnetism, which connects field strength to accretion rates. Within this radii, the measured accretion rates translate to $\sim2-3\times10^{-8}M_{\odot}$ yr$^{-1}$ based on Equation \eqref{eq:mdot_conv}.
In the wind driven accretion scenario,
we employ Equation (3) of \citet{wbf21}, applying $m\approx 4$ (ratio of toroidal field between midplane and wind base), $f'\approx2$ (ratio of $B_\phi$ and $B_z$ at the wind base), and $R=30$ AU,
we find a magnetic field strength of $\sim 0.008$ G, in good agreement with the actual measurements.
Both of these results are also consistent with constraints placed by paleomagnetic measurements for the Jupiter family comet 67P, expected to form in the region between 15 to 45 AU, of a local field strength $<0.03$ G \citep{biersteker19}.

\section{Conclusions}\label{sec:cc}

In this paper, we study the gas dynamics and magnetic field evolution in outer regions of PPDs. To this end, we conduct a set of global 3D non-ideal MHD simulations with ambipolar diffusion, achieving numerical resolutions comparable to those in local shearing box simulations. Our simulations are designed to be scale-free, and we perform parameter study on the strengths of AD and levels of disk magnetization. We summarize our main findings as follows.

\begin{enumerate}
    \item Our simulations demonstrate the co-existence of magnetized disk winds and MRI turbulence. While MHD winds dominate the disk angular momentum transport, MRI turbulence also makes considerable contributions. 
    \item For the fiducial model \textsc{Am1} with $Am=1$ and $\beta_0=10^4$,  the turbulence level is on the order of $\delta v\sim0.03c_s$, and the vertically-integrated  $\alpha$ parameter is $\sim10^{-3}$, dominated by the Maxwell stress. Turbulence and stress levels increase with weaker AD or stronger magnetization, achieving up to $\delta v\sim0.1c_s$, $\alpha\sim10^{-2}$ for Models \textsc{Am10E} and \textsc{B3}, and down to $\delta v\sim0.01c_s$, but still $\alpha\sim10^{-3}$ for Model \textsc{Am0.1}.
    \item Spontaneous concentration of poloidal magnetic flux into largely axisymmetric flux sheets are observed in all of our simulation models, leading to radial variations in turbulence levels, stresses, and mass accretion rates.  With angular momentum transport being inherently non-smooth due to magnetic flux concentration, annular substructures arise as a natural consequence.
    \item In the fiducial model, the flux sheets have typical thickness of $\sim 0.5H$, and are dynamically evolving. The widths of the resulting rings or gaps are $1.5-2.5H$ in surface density, with surface density variations span over $15-50\%$, while they show more diverse properties as magnetization and AD Els\"{a}sser number vary. 
    \item The disk generally loses magnetic flux over time, though flux sheets may serve to prevent the loss of magnetic flux, as in Model \textsc{B3} and Model \textsc{Am1}.

\end{enumerate}

Our results show that the contrast in surface density variation due to the magnetic flux concentration is on the order of unity or even less, and they evolve as flux sheets stochastically migrate radially in disks.
This implies that the magnetically-induced annular substructures are typically shallow, dynamically evolving, and show diverse properties. It may account for some of the observed disk substructures, such as young disks in Class 0/I phase, for which the evolutionary stage may be too early to form Jovian planets, as well as finer substructures found in super-resolution images of DSHARP disk samples \citep{jennings+21}.

In surveying a wide range of parameter space, we have chosen relative small azimuthal domain size to reduce computational cost. Larger domain size may be needed to fully capture the global properties of the MRI turbulence (e.g., \citealp{flock12,suzuki14}). It will also lower the threshold in surface density variation to trigger the RWI \citep{ono_etal16}, which may potentially drive vortex formation. In addition, more self-consistent treatment of ionization and thermodynamics/radiative-transfer is needed towards more realistic study and characterization of the gas dynamics in the outer PPDs. These extensions will be explored in future works.

\section*{Acknowledgements}

We are pleased to thank an anonymous referee for the very prompt response and useful comments that improved the presentation of this paper. We thank Henrik Latter and Gordon Ogilvie for fruitful discussions. CC acknowledges the support from Department of Applied Mathematics and Theoretical Physics at the University of Cambridge. XNB acknowledges the support from the National Key R\&D Program of China (No.2019YFA0405100). Numerical simulations are conducted on TianHe-1 (A) at National Supercomputer Center in Tianjin, China, and on the Orion cluster at Department of Astronomy, Tsinghua University.

\section*{Data Availability}

The data underlying this article will be shared on reasonable request to the corresponding author.

\bsp
\label{lastpage}

\bibliographystyle{mnras}
\bibliography{disk}

\begin{thebibliography}{}
\makeatletter
\relax
\def\mn@urlcharsother{\let\do\@makeother \do\$\do\&\do\#\do\^\do\_\do\%\do\~}
\def\mn@doi{\begingroup\mn@urlcharsother \@ifnextchar [ {\mn@doi@}
  {\mn@doi@[]}}
\def\mn@doi@[#1]#2{\def\@tempa{#1}\ifx\@tempa\@empty \href
  {http://dx.doi.org/#2} {doi:#2}\else \href {http://dx.doi.org/#2} {#1}\fi
  \endgroup}
\def\mn@eprint#1#2{\mn@eprint@#1:#2::\@nil}
\def\mn@eprint@arXiv#1{\href {http://arxiv.org/abs/#1} {{\tt arXiv:#1}}}
\def\mn@eprint@dblp#1{\href {http://dblp.uni-trier.de/rec/bibtex/#1.xml}
  {dblp:#1}}
\def\mn@eprint@#1:#2:#3:#4\@nil{\def\@tempa {#1}\def\@tempb {#2}\def\@tempc
  {#3}\ifx \@tempc \@empty \let \@tempc \@tempb \let \@tempb \@tempa \fi \ifx
  \@tempb \@empty \def\@tempb {arXiv}\fi \@ifundefined
  {mn@eprint@\@tempb}{\@tempb:\@tempc}{\expandafter \expandafter \csname
  mn@eprint@\@tempb\endcsname \expandafter{\@tempc}}}

\bibitem[\protect\citeauthoryear{{ALMA Partnership} et~al.,}{{ALMA Partnership}
  et~al.}{2015}]{alma_etal15}
{ALMA Partnership} et~al., 2015, \mn@doi [\apjl] {10.1088/2041-8205/808/1/L3},
  \href {https://ui.adsabs.harvard.edu/abs/2015ApJ...808L...3A} {808, L3}

\bibitem[\protect\citeauthoryear{{Alexiades}, {Amiez}  \&
  {Gremaud}}{{Alexiades} et~al.}{1996}]{alexiades96}
{Alexiades} V.,  {Amiez} G.,   {Gremaud} P.,  1996, Communications in Numerical
  Methods in Engineering, 12, 31

\bibitem[\protect\citeauthoryear{{Andrews}}{{Andrews}}{2020}]{andrews20}
{Andrews} S.~M.,  2020, arXiv e-prints, \href
  {https://ui.adsabs.harvard.edu/abs/2020arXiv200105007A} {p. arXiv:2001.05007}

\bibitem[\protect\citeauthoryear{{Andrews} et~al.,}{{Andrews}
  et~al.}{2016}]{andrews_etal16}
{Andrews} S.~M.,  et~al., 2016, \mn@doi [\apjl] {10.3847/2041-8205/820/2/L40},
  \href {https://ui.adsabs.harvard.edu/abs/2016ApJ...820L..40A} {820, L40}

\bibitem[\protect\citeauthoryear{{Andrews} et~al.,}{{Andrews}
  et~al.}{2018}]{andrews+18}
{Andrews} S.~M.,  et~al., 2018, \mn@doi [\apjl] {10.3847/2041-8213/aaf741},
  \href {https://ui.adsabs.harvard.edu/abs/2018ApJ...869L..41A} {869, L41}

\bibitem[\protect\citeauthoryear{{Armitage}}{{Armitage}}{2010}]{armitage10}
{Armitage} P.~J.,  2010, {Astrophysics of Planet Formation}

\bibitem[\protect\citeauthoryear{{Avenhaus} et~al.,}{{Avenhaus}
  et~al.}{2018}]{avenhaus_etal18}
{Avenhaus} H.,  et~al., 2018, \mn@doi [\apj] {10.3847/1538-4357/aab846}, \href
  {https://ui.adsabs.harvard.edu/abs/2018ApJ...863...44A} {863, 44}

\bibitem[\protect\citeauthoryear{{Bai}}{{Bai}}{2011a}]{bai11a}
{Bai} X.-N.,  2011a, \mn@doi [\apj] {10.1088/0004-637X/739/1/50}, \href
  {https://ui.adsabs.harvard.edu/abs/2011ApJ...739...50B} {739, 50}

\bibitem[\protect\citeauthoryear{{Bai}}{{Bai}}{2011b}]{bai11b}
{Bai} X.-N.,  2011b, \mn@doi [\apj] {10.1088/0004-637X/739/1/51}, \href
  {https://ui.adsabs.harvard.edu/abs/2011ApJ...739...51B} {739, 51}

\bibitem[\protect\citeauthoryear{{Bai}}{{Bai}}{2013}]{bai13}
{Bai} X.-N.,  2013, \mn@doi [\apj] {10.1088/0004-637X/772/2/96}, \href
  {https://ui.adsabs.harvard.edu/abs/2013ApJ...772...96B} {772, 96}

\bibitem[\protect\citeauthoryear{{Bai}}{{Bai}}{2014}]{bai14}
{Bai} X.-N.,  2014, \mn@doi [\apj] {10.1088/0004-637X/791/2/137}, \href
  {https://ui.adsabs.harvard.edu/abs/2014ApJ...791..137B} {791, 137}

\bibitem[\protect\citeauthoryear{{Bai}}{{Bai}}{2015}]{bai15}
{Bai} X.-N.,  2015, \mn@doi [\apj] {10.1088/0004-637X/798/2/84}, \href
  {https://ui.adsabs.harvard.edu/abs/2015ApJ...798...84B} {798, 84}

\bibitem[\protect\citeauthoryear{{Bai}}{{Bai}}{2017}]{bai17}
{Bai} X.-N.,  2017, \mn@doi [\apj] {10.3847/1538-4357/aa7dda}, \href
  {https://ui.adsabs.harvard.edu/abs/2017ApJ...845...75B} {845, 75}

\bibitem[\protect\citeauthoryear{{Bai} \& {Goodman}}{{Bai} \&
  {Goodman}}{2009}]{BG09}
{Bai} X.-N.,  {Goodman} J.,  2009, \mn@doi [\apj]
  {10.1088/0004-637X/701/1/737}, \href
  {https://ui.adsabs.harvard.edu/abs/2009ApJ...701..737B} {701, 737}

\bibitem[\protect\citeauthoryear{{Bai} \& {Stone}}{{Bai} \&
  {Stone}}{2011}]{bs11}
{Bai} X.-N.,  {Stone} J.~M.,  2011, \mn@doi [\apj]
  {10.1088/0004-637X/736/2/144}, \href
  {https://ui.adsabs.harvard.edu/abs/2011ApJ...736..144B} {736, 144}

\bibitem[\protect\citeauthoryear{{Bai} \& {Stone}}{{Bai} \&
  {Stone}}{2013a}]{bs13a}
{Bai} X.-N.,  {Stone} J.~M.,  2013a, \mn@doi [\apj]
  {10.1088/0004-637X/767/1/30}, \href
  {https://ui.adsabs.harvard.edu/abs/2013ApJ...767...30B} {767, 30}

\bibitem[\protect\citeauthoryear{{Bai} \& {Stone}}{{Bai} \&
  {Stone}}{2013b}]{bs13}
{Bai} X.-N.,  {Stone} J.~M.,  2013b, \mn@doi [\apj]
  {10.1088/0004-637X/769/1/76}, \href
  {https://ui.adsabs.harvard.edu/abs/2013ApJ...769...76B} {769, 76}

\bibitem[\protect\citeauthoryear{{Bai} \& {Stone}}{{Bai} \&
  {Stone}}{2014}]{bs14}
{Bai} X.-N.,  {Stone} J.~M.,  2014, \mn@doi [\apj]
  {10.1088/0004-637X/796/1/31}, \href
  {https://ui.adsabs.harvard.edu/abs/2014ApJ...796...31B} {796, 31}

\bibitem[\protect\citeauthoryear{{Bai} \& {Stone}}{{Bai} \&
  {Stone}}{2017}]{bs17}
{Bai} X.-N.,  {Stone} J.~M.,  2017, \mn@doi [\apj]
  {10.3847/1538-4357/836/1/46}, \href
  {https://ui.adsabs.harvard.edu/abs/2017ApJ...836...46B} {836, 46}

\bibitem[\protect\citeauthoryear{{Bai}, {Ye}, {Goodman}  \& {Yuan}}{{Bai}
  et~al.}{2016}]{bai+16}
{Bai} X.-N.,  {Ye} J.,  {Goodman} J.,   {Yuan} F.,  2016, \mn@doi [\apj]
  {10.3847/0004-637X/818/2/152}, \href
  {https://ui.adsabs.harvard.edu/abs/2016ApJ...818..152B} {818, 152}

\bibitem[\protect\citeauthoryear{{Balbus} \& {Hawley}}{{Balbus} \&
  {Hawley}}{1991}]{bh91}
{Balbus} S.~A.,  {Hawley} J.~F.,  1991, \mn@doi [\apj] {10.1086/170270}, \href
  {https://ui.adsabs.harvard.edu/abs/1991ApJ...376..214B} {376, 214}

\bibitem[\protect\citeauthoryear{{Benisty} et~al.,}{{Benisty}
  et~al.}{2015}]{benisty_etal15}
{Benisty} M.,  et~al., 2015, \mn@doi [\aap] {10.1051/0004-6361/201526011},
  \href {https://ui.adsabs.harvard.edu/abs/2015A&A...578L...6B} {578, L6}

\bibitem[\protect\citeauthoryear{{B{\'e}thune}, {Lesur}  \&
  {Ferreira}}{{B{\'e}thune} et~al.}{2016}]{bethune+16}
{B{\'e}thune} W.,  {Lesur} G.,   {Ferreira} J.,  2016, \mn@doi [\aap]
  {10.1051/0004-6361/201527874}, \href
  {https://ui.adsabs.harvard.edu/abs/2016A&A...589A..87B} {589, A87}

\bibitem[\protect\citeauthoryear{{B{\'e}thune}, {Lesur}  \&
  {Ferreira}}{{B{\'e}thune} et~al.}{2017}]{bethune_etal17}
{B{\'e}thune} W.,  {Lesur} G.,   {Ferreira} J.,  2017, \mn@doi [\aap]
  {10.1051/0004-6361/201630056}, \href
  {https://ui.adsabs.harvard.edu/abs/2017A&A...600A..75B} {600, A75}

\bibitem[\protect\citeauthoryear{{Biersteker}, {Weiss}, {Heinisch},
  {Her{\v{c}}ik}, {Glassmeier}  \& {Auster}}{{Biersteker}
  et~al.}{2019}]{biersteker19}
{Biersteker} J.~B.,  {Weiss} B.~P.,  {Heinisch} P.,  {Her{\v{c}}ik} D.,
  {Glassmeier} K.-H.,   {Auster} H.-U.,  2019, \mn@doi [\apj]
  {10.3847/1538-4357/ab0f2a}, \href
  {https://ui.adsabs.harvard.edu/abs/2019ApJ...875...39B} {875, 39}

\bibitem[\protect\citeauthoryear{{Blandford} \& {Payne}}{{Blandford} \&
  {Payne}}{1982}]{bp82}
{Blandford} R.~D.,  {Payne} D.~G.,  1982, \mn@doi [\mnras]
  {10.1093/mnras/199.4.883}, \href
  {https://ui.adsabs.harvard.edu/abs/1982MNRAS.199..883B} {199, 883}

\bibitem[\protect\citeauthoryear{{Brandenburg} \& {Zweibel}}{{Brandenburg} \&
  {Zweibel}}{1994}]{bz94}
{Brandenburg} A.,  {Zweibel} E.~G.,  1994, \mn@doi [\apjl] {10.1086/187372},
  \href {https://ui.adsabs.harvard.edu/abs/1994ApJ...427L..91B} {427, L91}

\bibitem[\protect\citeauthoryear{{Brauer}, {Wolf}  \& {Flock}}{{Brauer}
  et~al.}{2017}]{brauer_etal17}
{Brauer} R.,  {Wolf} S.,   {Flock} M.,  2017, \mn@doi [\aap]
  {10.1051/0004-6361/201731140}, \href
  {https://ui.adsabs.harvard.edu/abs/2017A&A...607A.104B} {607, A104}

\bibitem[\protect\citeauthoryear{{Bryden}, {Chen}, {Lin}, {Nelson}  \&
  {Papaloizou}}{{Bryden} et~al.}{1999}]{bryden_etal99}
{Bryden} G.,  {Chen} X.,  {Lin} D.~N.~C.,  {Nelson} R.~P.,   {Papaloizou} J.
  C.~B.,  1999, \mn@doi [\apj] {10.1086/306917}, \href
  {https://ui.adsabs.harvard.edu/abs/1999ApJ...514..344B} {514, 344}

\bibitem[\protect\citeauthoryear{{Cazzoletti}, {van Dishoeck}, {Visser},
  {Facchini}  \& {Bruderer}}{{Cazzoletti} et~al.}{2018}]{cazzoletti_etal18}
{Cazzoletti} P.,  {van Dishoeck} E.~F.,  {Visser} R.,  {Facchini} S.,
  {Bruderer} S.,  2018, \mn@doi [\aap] {10.1051/0004-6361/201731457}, \href
  {https://ui.adsabs.harvard.edu/abs/2018A&A...609A..93C} {609, A93}

\bibitem[\protect\citeauthoryear{{Chandrasekhar}}{{Chandrasekhar}}{1961}]{c61}
{Chandrasekhar} S.,  1961, {Hydrodynamic and hydromagnetic stability}

\bibitem[\protect\citeauthoryear{{Cui} \& {Bai}}{{Cui} \& {Bai}}{2020}]{cb20}
{Cui} C.,  {Bai} X.-N.,  2020, \mn@doi [\apj] {10.3847/1538-4357/ab7194}, \href
  {https://ui.adsabs.harvard.edu/abs/2020ApJ...891...30C} {891, 30}

\bibitem[\protect\citeauthoryear{{Cui} \& {Lin}}{{Cui} \& {Lin}}{2021}]{cl21}
{Cui} C.,  {Lin} M.-K.,  2021, \mn@doi [\mnras] {10.1093/mnras/stab1511}, \href
  {https://ui.adsabs.harvard.edu/abs/2021MNRAS.505.2983C} {505, 2983}

\bibitem[\protect\citeauthoryear{{Davis}, {Stone}  \& {Pessah}}{{Davis}
  et~al.}{2010}]{davis+10}
{Davis} S.~W.,  {Stone} J.~M.,   {Pessah} M.~E.,  2010, \mn@doi [\apj]
  {10.1088/0004-637X/713/1/52}, \href
  {https://ui.adsabs.harvard.edu/abs/2010ApJ...713...52D} {713, 52}

\bibitem[\protect\citeauthoryear{{Dittrich}, {Klahr}  \& {Johansen}}{{Dittrich}
  et~al.}{2013}]{dittrich_etal13}
{Dittrich} K.,  {Klahr} H.,   {Johansen} A.,  2013, \mn@doi [\apj]
  {10.1088/0004-637X/763/2/117}, \href
  {https://ui.adsabs.harvard.edu/abs/2013ApJ...763..117D} {763, 117}

\bibitem[\protect\citeauthoryear{{Dong} \& {Fung}}{{Dong} \&
  {Fung}}{2017}]{df17}
{Dong} R.,  {Fung} J.,  2017, \mn@doi [\apj] {10.3847/1538-4357/835/2/146},
  \href {https://ui.adsabs.harvard.edu/abs/2017ApJ...835..146D} {835, 146}

\bibitem[\protect\citeauthoryear{{Dong}, {Li}, {Chiang}  \& {Li}}{{Dong}
  et~al.}{2017}]{dong_etal17}
{Dong} R.,  {Li} S.,  {Chiang} E.,   {Li} H.,  2017, \mn@doi [\apj]
  {10.3847/1538-4357/aa72f2}, \href
  {https://ui.adsabs.harvard.edu/abs/2017ApJ...843..127D} {843, 127}

\bibitem[\protect\citeauthoryear{{Drazkowska}, {Alibert}  \&
  {Moore}}{{Drazkowska} et~al.}{2016}]{drazkowka_etal16}
{Drazkowska} J.,  {Alibert} Y.,   {Moore} B.,  2016, \mn@doi [\aap]
  {10.1051/0004-6361/201628983}, \href
  {https://ui.adsabs.harvard.edu/abs/2016A&A...594A.105D} {594, A105}

\bibitem[\protect\citeauthoryear{{Dullemond} \& {Penzlin}}{{Dullemond} \&
  {Penzlin}}{2018}]{dp18}
{Dullemond} C.~P.,  {Penzlin} A.~B.~T.,  2018, \mn@doi [\aap]
  {10.1051/0004-6361/201731878}, \href
  {https://ui.adsabs.harvard.edu/abs/2018A&A...609A..50D} {609, A50}

\bibitem[\protect\citeauthoryear{{Dullemond} et~al.,}{{Dullemond}
  et~al.}{2018}]{dullemond+18}
{Dullemond} C.~P.,  et~al., 2018, \mn@doi [\apjl] {10.3847/2041-8213/aaf742},
  \href {https://ui.adsabs.harvard.edu/abs/2018ApJ...869L..46D} {869, L46}

\bibitem[\protect\citeauthoryear{{Flaherty} et~al.,}{{Flaherty}
  et~al.}{2017}]{flaherty_etal17}
{Flaherty} K.~M.,  et~al., 2017, \mn@doi [\apj] {10.3847/1538-4357/aa79f9},
  \href {https://ui.adsabs.harvard.edu/abs/2017ApJ...843..150F} {843, 150}

\bibitem[\protect\citeauthoryear{{Fleming} \& {Stone}}{{Fleming} \&
  {Stone}}{2003}]{fs03}
{Fleming} T.,  {Stone} J.~M.,  2003, \mn@doi [\apj] {10.1086/345848}, \href
  {https://ui.adsabs.harvard.edu/abs/2003ApJ...585..908F} {585, 908}

\bibitem[\protect\citeauthoryear{{Flock}, {Dzyurkevich}, {Klahr}, {Turner}  \&
  {Henning}}{{Flock} et~al.}{2011}]{flock+11}
{Flock} M.,  {Dzyurkevich} N.,  {Klahr} H.,  {Turner} N.~J.,   {Henning} T.,
  2011, \mn@doi [\apj] {10.1088/0004-637X/735/2/122}, \href
  {https://ui.adsabs.harvard.edu/abs/2011ApJ...735..122F} {735, 122}

\bibitem[\protect\citeauthoryear{{Flock}, {Dzyurkevich}, {Klahr}, {Turner}  \&
  {Henning}}{{Flock} et~al.}{2012}]{flock12}
{Flock} M.,  {Dzyurkevich} N.,  {Klahr} H.,  {Turner} N.,   {Henning} T.,
  2012, \mn@doi [\apj] {10.1088/0004-637X/744/2/144}, \href
  {https://ui.adsabs.harvard.edu/abs/2012ApJ...744..144F} {744, 144}

\bibitem[\protect\citeauthoryear{{Flock}, {Ruge}, {Dzyurkevich}, {Henning},
  {Klahr}  \& {Wolf}}{{Flock} et~al.}{2015}]{flock_etal15}
{Flock} M.,  {Ruge} J.~P.,  {Dzyurkevich} N.,  {Henning} T.,  {Klahr} H.,
  {Wolf} S.,  2015, \mn@doi [\aap] {10.1051/0004-6361/201424693}, \href
  {https://ui.adsabs.harvard.edu/abs/2015A&A...574A..68F} {574, A68}

\bibitem[\protect\citeauthoryear{{Fromang}, {Latter}, {Lesur}  \&
  {Ogilvie}}{{Fromang} et~al.}{2013}]{fromang_etal13}
{Fromang} S.,  {Latter} H.,  {Lesur} G.,   {Ogilvie} G.~I.,  2013, \mn@doi
  [\aap] {10.1051/0004-6361/201220016}, \href
  {https://ui.adsabs.harvard.edu/abs/2013A&A...552A..71F} {552, A71}

\bibitem[\protect\citeauthoryear{{Gammie}}{{Gammie}}{1996}]{gammie96}
{Gammie} C.~F.,  1996, \mn@doi [\apj] {10.1086/176735}, \href
  {https://ui.adsabs.harvard.edu/abs/1996ApJ...457..355G} {457, 355}

\bibitem[\protect\citeauthoryear{{Gardiner} \& {Stone}}{{Gardiner} \&
  {Stone}}{2005}]{gs05}
{Gardiner} T.~A.,  {Stone} J.~M.,  2005, \mn@doi [Journal of Computational
  Physics] {10.1016/j.jcp.2004.11.016}, \href
  {https://ui.adsabs.harvard.edu/abs/2005JCoPh.205..509G} {205, 509}

\bibitem[\protect\citeauthoryear{{Gardiner} \& {Stone}}{{Gardiner} \&
  {Stone}}{2008}]{gs08}
{Gardiner} T.~A.,  {Stone} J.~M.,  2008, \mn@doi [Journal of Computational
  Physics] {10.1016/j.jcp.2007.12.017}, \href
  {https://ui.adsabs.harvard.edu/abs/2008JCoPh.227.4123G} {227, 4123}

\bibitem[\protect\citeauthoryear{{Glassgold}, {Najita}  \& {Igea}}{{Glassgold}
  et~al.}{2004}]{glassgold_etal04}
{Glassgold} A.~E.,  {Najita} J.,   {Igea} J.,  2004, \mn@doi [\apj]
  {10.1086/424509}, \href
  {https://ui.adsabs.harvard.edu/abs/2004ApJ...615..972G} {615, 972}

\bibitem[\protect\citeauthoryear{{Goldreich} \& {Tremaine}}{{Goldreich} \&
  {Tremaine}}{1979}]{gt79}
{Goldreich} P.,  {Tremaine} S.,  1979, \mn@doi [\apj] {10.1086/157448}, \href
  {https://ui.adsabs.harvard.edu/abs/1979ApJ...233..857G} {233, 857}

\bibitem[\protect\citeauthoryear{{Gonzalez}, {Laibe}  \& {Maddison}}{{Gonzalez}
  et~al.}{2017}]{gonzalez_etal17}
{Gonzalez} J.~F.,  {Laibe} G.,   {Maddison} S.~T.,  2017, \mn@doi [\mnras]
  {10.1093/mnras/stx016}, \href
  {https://ui.adsabs.harvard.edu/abs/2017MNRAS.467.1984G} {467, 1984}

\bibitem[\protect\citeauthoryear{{Gressel}, {Turner}, {Nelson}  \&
  {McNally}}{{Gressel} et~al.}{2015}]{gressel_etal15}
{Gressel} O.,  {Turner} N.~J.,  {Nelson} R.~P.,   {McNally} C.~P.,  2015,
  \mn@doi [\apj] {10.1088/0004-637X/801/2/84}, \href
  {https://ui.adsabs.harvard.edu/abs/2015ApJ...801...84G} {801, 84}

\bibitem[\protect\citeauthoryear{{Gressel}, {Ramsey}, {Brinch}, {Nelson},
  {Turner}  \& {Bruderer}}{{Gressel} et~al.}{2020}]{gressel_etal20}
{Gressel} O.,  {Ramsey} J.~P.,  {Brinch} C.,  {Nelson} R.~P.,  {Turner} N.~J.,
   {Bruderer} S.,  2020, \mn@doi [\apj] {10.3847/1538-4357/ab91b7}, \href
  {https://ui.adsabs.harvard.edu/abs/2020ApJ...896..126G} {896, 126}

\bibitem[\protect\citeauthoryear{{Haisch}, {Lada}  \& {Lada}}{{Haisch}
  et~al.}{2001}]{haisch_etal01}
{Haisch} Karl~E. J.,  {Lada} E.~A.,   {Lada} C.~J.,  2001, \mn@doi [\apjl]
  {10.1086/320685}, \href
  {https://ui.adsabs.harvard.edu/abs/2001ApJ...553L.153H} {553, L153}

\bibitem[\protect\citeauthoryear{{Harrison} et~al.,}{{Harrison}
  et~al.}{2021}]{harrison_etal21}
{Harrison} R.~E.,  et~al., 2021, \mn@doi [\apj] {10.3847/1538-4357/abd94e},
  \href {https://ui.adsabs.harvard.edu/abs/2021ApJ...908..141H} {908, 141}

\bibitem[\protect\citeauthoryear{{Hartmann}, {Calvet}, {Gullbring}  \&
  {D'Alessio}}{{Hartmann} et~al.}{1998}]{hartmann_etal98}
{Hartmann} L.,  {Calvet} N.,  {Gullbring} E.,   {D'Alessio} P.,  1998, \mn@doi
  [\apj] {10.1086/305277}, \href
  {https://ui.adsabs.harvard.edu/abs/1998ApJ...495..385H} {495, 385}

\bibitem[\protect\citeauthoryear{{Hawley}, {Guan}  \& {Krolik}}{{Hawley}
  et~al.}{2011}]{hawley+11}
{Hawley} J.~F.,  {Guan} X.,   {Krolik} J.~H.,  2011, \mn@doi [\apj]
  {10.1088/0004-637X/738/1/84}, \href
  {https://ui.adsabs.harvard.edu/abs/2011ApJ...738...84H} {738, 84}

\bibitem[\protect\citeauthoryear{{Herczeg} \& {Hillenbrand}}{{Herczeg} \&
  {Hillenbrand}}{2008}]{hh08}
{Herczeg} G.~J.,  {Hillenbrand} L.~A.,  2008, \mn@doi [\apj] {10.1086/586728},
  \href {https://ui.adsabs.harvard.edu/abs/2008ApJ...681..594H} {681, 594}

\bibitem[\protect\citeauthoryear{{Hu} \& {Bai}}{{Hu} \& {Bai}}{2021}]{hb21}
{Hu} Z.,  {Bai} X.-N.,  2021, \mn@doi [\mnras] {10.1093/mnras/stab542}, \href
  {https://ui.adsabs.harvard.edu/abs/2021MNRAS.503..162H} {503, 162}

\bibitem[\protect\citeauthoryear{{Hu}, {Zhu}, {Okuzumi}, {Bai}, {Wang},
  {Tomida}  \& {Stone}}{{Hu} et~al.}{2019}]{hu_etal19}
{Hu} X.,  {Zhu} Z.,  {Okuzumi} S.,  {Bai} X.-N.,  {Wang} L.,  {Tomida} K.,
  {Stone} J.~M.,  2019, \mn@doi [\apj] {10.3847/1538-4357/ab44cb}, \href
  {https://ui.adsabs.harvard.edu/abs/2019ApJ...885...36H} {885, 36}

\bibitem[\protect\citeauthoryear{{Huang} et~al.,}{{Huang}
  et~al.}{2018a}]{huang_etal18}
{Huang} J.,  et~al., 2018a, \mn@doi [\apjl] {10.3847/2041-8213/aaf740}, \href
  {https://ui.adsabs.harvard.edu/abs/2018ApJ...869L..42H} {869, L42}

\bibitem[\protect\citeauthoryear{{Huang} et~al.,}{{Huang}
  et~al.}{2018b}]{huang_etal18b}
{Huang} J.,  et~al., 2018b, \mn@doi [\apjl] {10.3847/2041-8213/aaf7a0}, \href
  {https://ui.adsabs.harvard.edu/abs/2018ApJ...869L..43H} {869, L43}

\bibitem[\protect\citeauthoryear{{Ilgner} \& {Nelson}}{{Ilgner} \&
  {Nelson}}{2008}]{ilgnernelson08}
{Ilgner} M.,  {Nelson} R.~P.,  2008, \mn@doi [\aap]
  {10.1051/0004-6361:20079307}, \href
  {https://ui.adsabs.harvard.edu/abs/2008A&A...483..815I} {483, 815}

\bibitem[\protect\citeauthoryear{{Isella} et~al.,}{{Isella}
  et~al.}{2016}]{isella_etal16}
{Isella} A.,  et~al., 2016, \mn@doi [\prl] {10.1103/PhysRevLett.117.251101},
  \href {https://ui.adsabs.harvard.edu/abs/2016PhRvL.117y1101I} {117, 251101}

\bibitem[\protect\citeauthoryear{{Jennings}, {Booth}, {Tazzari}, {Clarke}  \&
  {Rosotti}}{{Jennings} et~al.}{2021}]{jennings+21}
{Jennings} J.,  {Booth} R.~A.,  {Tazzari} M.,  {Clarke} C.~J.,   {Rosotti}
  G.~P.,  2021, arXiv e-prints, \href
  {https://ui.adsabs.harvard.edu/abs/2021arXiv210302392J} {p. arXiv:2103.02392}

\bibitem[\protect\citeauthoryear{{Johansen}, {Youdin}  \& {Klahr}}{{Johansen}
  et~al.}{2009}]{johansen_etal09}
{Johansen} A.,  {Youdin} A.,   {Klahr} H.,  2009, \mn@doi [\apj]
  {10.1088/0004-637X/697/2/1269}, \href
  {https://ui.adsabs.harvard.edu/abs/2009ApJ...697.1269J} {697, 1269}

\bibitem[\protect\citeauthoryear{{Klahr} \& {Bodenheimer}}{{Klahr} \&
  {Bodenheimer}}{2003}]{kb03}
{Klahr} H.~H.,  {Bodenheimer} P.,  2003, \mn@doi [\apj] {10.1086/344743}, \href
  {https://ui.adsabs.harvard.edu/abs/2003ApJ...582..869K} {582, 869}

\bibitem[\protect\citeauthoryear{{Klahr} \& {Hubbard}}{{Klahr} \&
  {Hubbard}}{2014}]{kh14}
{Klahr} H.,  {Hubbard} A.,  2014, \mn@doi [\apj] {10.1088/0004-637X/788/1/21},
  \href {https://ui.adsabs.harvard.edu/abs/2014ApJ...788...21K} {788, 21}

\bibitem[\protect\citeauthoryear{{Kley} \& {Nelson}}{{Kley} \&
  {Nelson}}{2012}]{kn12}
{Kley} W.,  {Nelson} R.~P.,  2012, \mn@doi [\araa]
  {10.1146/annurev-astro-081811-125523}, \href
  {https://ui.adsabs.harvard.edu/abs/2012ARA&A..50..211K} {50, 211}

\bibitem[\protect\citeauthoryear{{Krapp}, {Gressel}, {Ben{\'\i}tez-Llambay},
  {Downes}, {Mohandas}  \& {Pessah}}{{Krapp} et~al.}{2018}]{krapp+18}
{Krapp} L.,  {Gressel} O.,  {Ben{\'\i}tez-Llambay} P.,  {Downes} T.~P.,
  {Mohandas} G.,   {Pessah} M.~E.,  2018, \mn@doi [\apj]
  {10.3847/1538-4357/aadcf0}, \href
  {https://ui.adsabs.harvard.edu/abs/2018ApJ...865..105K} {865, 105}

\bibitem[\protect\citeauthoryear{{Kunz}}{{Kunz}}{2008}]{kunz08}
{Kunz} M.~W.,  2008, \mn@doi [\mnras] {10.1111/j.1365-2966.2008.12928.x}, \href
  {https://ui.adsabs.harvard.edu/abs/2008MNRAS.385.1494K} {385, 1494}

\bibitem[\protect\citeauthoryear{{Kunz} \& {Lesur}}{{Kunz} \&
  {Lesur}}{2013}]{kunz+13}
{Kunz} M.~W.,  {Lesur} G.,  2013, \mn@doi [\mnras] {10.1093/mnras/stt1171},
  \href {https://ui.adsabs.harvard.edu/abs/2013MNRAS.434.2295K} {434, 2295}

\bibitem[\protect\citeauthoryear{{Lesur}, {Kunz}  \& {Fromang}}{{Lesur}
  et~al.}{2014}]{lesur_etal14}
{Lesur} G.,  {Kunz} M.~W.,   {Fromang} S.,  2014, \mn@doi [\aap]
  {10.1051/0004-6361/201423660}, \href
  {https://ui.adsabs.harvard.edu/abs/2014A&A...566A..56L} {566, A56}

\bibitem[\protect\citeauthoryear{{Li}, {Finn}, {Lovelace}  \& {Colgate}}{{Li}
  et~al.}{2000}]{li_etal00}
{Li} H.,  {Finn} J.~M.,  {Lovelace} R.~V.~E.,   {Colgate} S.~A.,  2000, \mn@doi
  [\apj] {10.1086/308693}, \href
  {https://ui.adsabs.harvard.edu/abs/2000ApJ...533.1023L} {533, 1023}

\bibitem[\protect\citeauthoryear{{Lin} \& {Papaloizou}}{{Lin} \&
  {Papaloizou}}{1986}]{lp86}
{Lin} D.~N.~C.,  {Papaloizou} J.,  1986, \mn@doi [\apj] {10.1086/164653}, \href
  {https://ui.adsabs.harvard.edu/abs/1986ApJ...309..846L} {309, 846}

\bibitem[\protect\citeauthoryear{{Lin} \& {Papaloizou}}{{Lin} \&
  {Papaloizou}}{1993}]{lp93}
{Lin} D.~N.~C.,  {Papaloizou} J.~C.~B.,  1993, in {Levy} E.~H.,  {Lunine}
  J.~I.,  eds, Protostars and Planets III. p.~749

\bibitem[\protect\citeauthoryear{{Lin} \& {Youdin}}{{Lin} \&
  {Youdin}}{2015}]{ly15}
{Lin} M.-K.,  {Youdin} A.~N.,  2015, \mn@doi [\apj]
  {10.1088/0004-637X/811/1/17}, \href
  {https://ui.adsabs.harvard.edu/abs/2015ApJ...811...17L} {811, 17}

\bibitem[\protect\citeauthoryear{{Long} et~al.,}{{Long}
  et~al.}{2018}]{long_etal18}
{Long} F.,  et~al., 2018, \mn@doi [\apj] {10.3847/1538-4357/aae8e1}, \href
  {https://ui.adsabs.harvard.edu/abs/2018ApJ...869...17L} {869, 17}

\bibitem[\protect\citeauthoryear{{Lovelace}, {Li}, {Colgate}  \&
  {Nelson}}{{Lovelace} et~al.}{1999}]{lovelace99}
{Lovelace} R.~V.~E.,  {Li} H.,  {Colgate} S.~A.,   {Nelson} A.~F.,  1999,
  \mn@doi [\apj] {10.1086/306900}, \href
  {https://ui.adsabs.harvard.edu/abs/1999ApJ...513..805L} {513, 805}

\bibitem[\protect\citeauthoryear{{Lyra} \& {Umurhan}}{{Lyra} \&
  {Umurhan}}{2019}]{lu19}
{Lyra} W.,  {Umurhan} O.~M.,  2019, \mn@doi [\pasp] {10.1088/1538-3873/aaf5ff},
  \href {https://ui.adsabs.harvard.edu/abs/2019PASP..131g2001L} {131, 072001}

\bibitem[\protect\citeauthoryear{{Mamajek}}{{Mamajek}}{2009}]{mamajek09}
{Mamajek} E.~E.,  2009, in {Usuda} T.,  {Tamura} M.,   {Ishii} M.,  eds,
  American Institute of Physics Conference Series Vol. 1158, Exoplanets and
  Disks: Their Formation and Diversity. pp 3--10 (\mn@eprint {arXiv}
  {0906.5011}), \mn@doi{10.1063/1.3215910}

\bibitem[\protect\citeauthoryear{{Marcus}, {Pei}, {Jiang}  \&
  {Hassanzadeh}}{{Marcus} et~al.}{2013}]{marcus_etal13}
{Marcus} P.~S.,  {Pei} S.,  {Jiang} C.-H.,   {Hassanzadeh} P.,  2013, \mn@doi
  [\prl] {10.1103/PhysRevLett.111.084501}, \href
  {https://ui.adsabs.harvard.edu/abs/2013PhRvL.111h4501M} {111, 084501}

\bibitem[\protect\citeauthoryear{{Mazzei}, {Cleeves}  \& {Li}}{{Mazzei}
  et~al.}{2020}]{mazzei_etal20}
{Mazzei} R.,  {Cleeves} L.~I.,   {Li} Z.-Y.,  2020, \mn@doi [\apj]
  {10.3847/1538-4357/abb67a}, \href
  {https://ui.adsabs.harvard.edu/abs/2020ApJ...903...20M} {903, 20}

\bibitem[\protect\citeauthoryear{{Nakatani}, {Liu}, {Ohashi}, {Zhang},
  {Hanawa}, {Chandler}, {Oya}  \& {Sakai}}{{Nakatani}
  et~al.}{2020}]{nakatani_etal20}
{Nakatani} R.,  {Liu} H.~B.,  {Ohashi} S.,  {Zhang} Y.,  {Hanawa} T.,
  {Chandler} C.,  {Oya} Y.,   {Sakai} N.,  2020, \mn@doi [\apjl]
  {10.3847/2041-8213/ab8eaa}, \href
  {https://ui.adsabs.harvard.edu/abs/2020ApJ...895L...2N} {895, L2}

\bibitem[\protect\citeauthoryear{{Nelson}, {Papaloizou}, {Masset}  \&
  {Kley}}{{Nelson} et~al.}{2000}]{nelson_etal00}
{Nelson} R.~P.,  {Papaloizou} J. C.~B.,  {Masset} F.,   {Kley} W.,  2000,
  \mn@doi [\mnras] {10.1046/j.1365-8711.2000.03605.x}, \href
  {https://ui.adsabs.harvard.edu/abs/2000MNRAS.318...18N} {318, 18}

\bibitem[\protect\citeauthoryear{{Nelson}, {Gressel}  \& {Umurhan}}{{Nelson}
  et~al.}{2013}]{nelson_etal13}
{Nelson} R.~P.,  {Gressel} O.,   {Umurhan} O.~M.,  2013, \mn@doi [\mnras]
  {10.1093/mnras/stt1475}, \href
  {https://ui.adsabs.harvard.edu/abs/2013MNRAS.435.2610N} {435, 2610}

\bibitem[\protect\citeauthoryear{{Oishi} \& {Mac Low}}{{Oishi} \& {Mac
  Low}}{2009}]{oishi_etal09}
{Oishi} J.~S.,  {Mac Low} M.-M.,  2009, \mn@doi [\apj]
  {10.1088/0004-637X/704/2/1239}, \href
  {https://ui.adsabs.harvard.edu/abs/2009ApJ...704.1239O} {704, 1239}

\bibitem[\protect\citeauthoryear{{Okuzumi}, {Momose}, {Sirono}, {Kobayashi}  \&
  {Tanaka}}{{Okuzumi} et~al.}{2016}]{okuzuni_etal16}
{Okuzumi} S.,  {Momose} M.,  {Sirono} S.-i.,  {Kobayashi} H.,   {Tanaka} H.,
  2016, \mn@doi [\apj] {10.3847/0004-637X/821/2/82}, \href
  {https://ui.adsabs.harvard.edu/abs/2016ApJ...821...82O} {821, 82}

\bibitem[\protect\citeauthoryear{{Ono}, {Muto}, {Takeuchi}  \& {Nomura}}{{Ono}
  et~al.}{2016}]{ono_etal16}
{Ono} T.,  {Muto} T.,  {Takeuchi} T.,   {Nomura} H.,  2016, \mn@doi [\apj]
  {10.3847/0004-637X/823/2/84}, \href
  {https://ui.adsabs.harvard.edu/abs/2016ApJ...823...84O} {823, 84}

\bibitem[\protect\citeauthoryear{{Owen}}{{Owen}}{2020}]{owen20}
{Owen} J.~E.,  2020, \mn@doi [\mnras] {10.1093/mnras/staa1309}, \href
  {https://ui.adsabs.harvard.edu/abs/2020MNRAS.495.3160O} {495, 3160}

\bibitem[\protect\citeauthoryear{{Pandey} \& {Wardle}}{{Pandey} \&
  {Wardle}}{2012}]{pw12}
{Pandey} B.~P.,  {Wardle} M.,  2012, \mn@doi [\mnras]
  {10.1111/j.1365-2966.2012.20799.x}, \href
  {https://ui.adsabs.harvard.edu/abs/2012MNRAS.423..222P} {423, 222}

\bibitem[\protect\citeauthoryear{{Perez-Becker} \& {Chiang}}{{Perez-Becker} \&
  {Chiang}}{2011}]{pc11}
{Perez-Becker} D.,  {Chiang} E.,  2011, \mn@doi [\apj]
  {10.1088/0004-637X/735/1/8}, \href
  {https://ui.adsabs.harvard.edu/abs/2011ApJ...735....8P} {735, 8}

\bibitem[\protect\citeauthoryear{{P{\'e}rez} et~al.,}{{P{\'e}rez}
  et~al.}{2016}]{perez_etal16}
{P{\'e}rez} L.~M.,  et~al., 2016, \mn@doi [Science] {10.1126/science.aaf8296},
  \href {https://ui.adsabs.harvard.edu/abs/2016Sci...353.1519P} {353, 1519}

\bibitem[\protect\citeauthoryear{{Pinte} et~al.,}{{Pinte}
  et~al.}{2019}]{pinte_etal19}
{Pinte} C.,  et~al., 2019, \mn@doi [Nature Astronomy]
  {10.1038/s41550-019-0852-6}, \href
  {https://ui.adsabs.harvard.edu/abs/2019NatAs...3.1109P} {3, 1109}

\bibitem[\protect\citeauthoryear{{Pinte} et~al.,}{{Pinte}
  et~al.}{2020}]{pinte_etal20}
{Pinte} C.,  et~al., 2020, \mn@doi [\apjl] {10.3847/2041-8213/ab6dda}, \href
  {https://ui.adsabs.harvard.edu/abs/2020ApJ...890L...9P} {890, L9}

\bibitem[\protect\citeauthoryear{{Riols} \& {Lesur}}{{Riols} \&
  {Lesur}}{2018}]{rl18}
{Riols} A.,  {Lesur} G.,  2018, \mn@doi [\aap] {10.1051/0004-6361/201833212},
  \href {https://ui.adsabs.harvard.edu/abs/2018A&A...617A.117R} {617, A117}

\bibitem[\protect\citeauthoryear{{Riols} \& {Lesur}}{{Riols} \&
  {Lesur}}{2019}]{rl19}
{Riols} A.,  {Lesur} G.,  2019, \mn@doi [\aap] {10.1051/0004-6361/201834813},
  \href {https://ui.adsabs.harvard.edu/abs/2019A&A...625A.108R} {625, A108}

\bibitem[\protect\citeauthoryear{{Riols}, {Ogilvie}, {Latter}  \&
  {Ross}}{{Riols} et~al.}{2016}]{riols16}
{Riols} A.,  {Ogilvie} G.~I.,  {Latter} H.,   {Ross} J.~P.,  2016, \mn@doi
  [\mnras] {10.1093/mnras/stw2196}, \href
  {https://ui.adsabs.harvard.edu/abs/2016MNRAS.463.3096R} {463, 3096}

\bibitem[\protect\citeauthoryear{{Riols}, {Lesur}  \& {Menard}}{{Riols}
  et~al.}{2020}]{rl20}
{Riols} A.,  {Lesur} G.,   {Menard} F.,  2020, \mn@doi [\aap]
  {10.1051/0004-6361/201937418}, \href
  {https://ui.adsabs.harvard.edu/abs/2020A&A...639A..95R} {639, A95}

\bibitem[\protect\citeauthoryear{{Rodenkirch}, {Klahr}, {Fendt}  \&
  {Dullemond}}{{Rodenkirch} et~al.}{2020}]{rodenkirch20}
{Rodenkirch} P.~J.,  {Klahr} H.,  {Fendt} C.,   {Dullemond} C.~P.,  2020,
  \mn@doi [\aap] {10.1051/0004-6361/201834945}, \href
  {https://ui.adsabs.harvard.edu/abs/2020A&A...633A..21R} {633, A21}

\bibitem[\protect\citeauthoryear{{Rosotti}, {Teague}, {Dullemond}, {Booth}  \&
  {Clarke}}{{Rosotti} et~al.}{2020}]{rosotti+20}
{Rosotti} G.~P.,  {Teague} R.,  {Dullemond} C.,  {Booth} R.~A.,   {Clarke}
  C.~J.,  2020, \mn@doi [\mnras] {10.1093/mnras/staa1170}, \href
  {https://ui.adsabs.harvard.edu/abs/2020MNRAS.495..173R} {495, 173}

\bibitem[\protect\citeauthoryear{{Segura-Cox} et~al.,}{{Segura-Cox}
  et~al.}{2020}]{segura-cox_etal20}
{Segura-Cox} D.~M.,  et~al., 2020, arXiv e-prints, \href
  {https://ui.adsabs.harvard.edu/abs/2020arXiv201003657S} {p. arXiv:2010.03657}

\bibitem[\protect\citeauthoryear{{Shakura} \& {Sunyaev}}{{Shakura} \&
  {Sunyaev}}{1973}]{ss73}
{Shakura} N.~I.,  {Sunyaev} R.~A.,  1973, \aap, \href
  {https://ui.adsabs.harvard.edu/abs/1973A&A....24..337S} {500, 33}

\bibitem[\protect\citeauthoryear{{Shariff} \& {Cuzzi}}{{Shariff} \&
  {Cuzzi}}{2011}]{sc11}
{Shariff} K.,  {Cuzzi} J.~N.,  2011, \mn@doi [\apj]
  {10.1088/0004-637X/738/1/73}, \href
  {https://ui.adsabs.harvard.edu/abs/2011ApJ...738...73S} {738, 73}

\bibitem[\protect\citeauthoryear{{Sheehan} \& {Eisner}}{{Sheehan} \&
  {Eisner}}{2018}]{se18}
{Sheehan} P.~D.,  {Eisner} J.~A.,  2018, \mn@doi [\apj]
  {10.3847/1538-4357/aaae65}, \href
  {https://ui.adsabs.harvard.edu/abs/2018ApJ...857...18S} {857, 18}

\bibitem[\protect\citeauthoryear{{Sheehan}, {Tobin}, {Federman}, {Megeath}  \&
  {Looney}}{{Sheehan} et~al.}{2020}]{sheehan_etal20}
{Sheehan} P.~D.,  {Tobin} J.~J.,  {Federman} S.,  {Megeath} S.~T.,   {Looney}
  L.~W.,  2020, \mn@doi [\apj] {10.3847/1538-4357/abbad5}, \href
  {https://ui.adsabs.harvard.edu/abs/2020ApJ...902..141S} {902, 141}

\bibitem[\protect\citeauthoryear{{Shi}, {Krolik}  \& {Hirose}}{{Shi}
  et~al.}{2010}]{shi+10}
{Shi} J.,  {Krolik} J.~H.,   {Hirose} S.,  2010, \mn@doi [\apj]
  {10.1088/0004-637X/708/2/1716}, \href
  {https://ui.adsabs.harvard.edu/abs/2010ApJ...708.1716S} {708, 1716}

\bibitem[\protect\citeauthoryear{{Simon} \& {Armitage}}{{Simon} \&
  {Armitage}}{2014}]{sa14}
{Simon} J.~B.,  {Armitage} P.~J.,  2014, \mn@doi [\apj]
  {10.1088/0004-637X/784/1/15}, \href
  {https://ui.adsabs.harvard.edu/abs/2014ApJ...784...15S} {784, 15}

\bibitem[\protect\citeauthoryear{{Simon}, {Bai}, {Stone}, {Armitage}  \&
  {Beckwith}}{{Simon} et~al.}{2013a}]{simon_etal13a}
{Simon} J.~B.,  {Bai} X.-N.,  {Stone} J.~M.,  {Armitage} P.~J.,   {Beckwith}
  K.,  2013a, \mn@doi [\apj] {10.1088/0004-637X/764/1/66}, \href
  {https://ui.adsabs.harvard.edu/abs/2013ApJ...764...66S} {764, 66}

\bibitem[\protect\citeauthoryear{{Simon}, {Bai}, {Armitage}, {Stone}  \&
  {Beckwith}}{{Simon} et~al.}{2013b}]{simon_etal13b}
{Simon} J.~B.,  {Bai} X.-N.,  {Armitage} P.~J.,  {Stone} J.~M.,   {Beckwith}
  K.,  2013b, \mn@doi [\apj] {10.1088/0004-637X/775/1/73}, \href
  {https://ui.adsabs.harvard.edu/abs/2013ApJ...775...73S} {775, 73}

\bibitem[\protect\citeauthoryear{{Simon}, {Bai}, {Flaherty}  \&
  {Hughes}}{{Simon} et~al.}{2018}]{simon_etal18}
{Simon} J.~B.,  {Bai} X.-N.,  {Flaherty} K.~M.,   {Hughes} A.~M.,  2018,
  \mn@doi [\apj] {10.3847/1538-4357/aad86d}, \href
  {https://ui.adsabs.harvard.edu/abs/2018ApJ...865...10S} {865, 10}

\bibitem[\protect\citeauthoryear{{Stone}, {Gardiner}, {Teuben}, {Hawley}  \&
  {Simon}}{{Stone} et~al.}{2008}]{stone_etal08}
{Stone} J.~M.,  {Gardiner} T.~A.,  {Teuben} P.,  {Hawley} J.~F.,   {Simon}
  J.~B.,  2008, \mn@doi [\apjs] {10.1086/588755}, \href
  {https://ui.adsabs.harvard.edu/abs/2008ApJS..178..137S} {178, 137}

\bibitem[\protect\citeauthoryear{{Stone}, {Tomida}, {White}  \&
  {Felker}}{{Stone} et~al.}{2020}]{stone_etal20}
{Stone} J.~M.,  {Tomida} K.,  {White} C.~J.,   {Felker} K.~G.,  2020, arXiv
  e-prints, \href {https://ui.adsabs.harvard.edu/abs/2020arXiv200506651S} {p.
  arXiv:2005.06651}

\bibitem[\protect\citeauthoryear{{Suriano}, {Li}, {Krasnopolsky}  \&
  {Shang}}{{Suriano} et~al.}{2018}]{suriano_etal18}
{Suriano} S.~S.,  {Li} Z.-Y.,  {Krasnopolsky} R.,   {Shang} H.,  2018, \mn@doi
  [\mnras] {10.1093/mnras/sty717}, \href
  {https://ui.adsabs.harvard.edu/abs/2018MNRAS.477.1239S} {477, 1239}

\bibitem[\protect\citeauthoryear{{Suriano}, {Li}, {Krasnopolsky}, {Suzuki}  \&
  {Shang}}{{Suriano} et~al.}{2019}]{suriano_etal19}
{Suriano} S.~S.,  {Li} Z.-Y.,  {Krasnopolsky} R.,  {Suzuki} T.~K.,   {Shang}
  H.,  2019, \mn@doi [\mnras] {10.1093/mnras/sty3502}, \href
  {https://ui.adsabs.harvard.edu/abs/2019MNRAS.484..107S} {484, 107}

\bibitem[\protect\citeauthoryear{{Suzuki} \& {Inutsuka}}{{Suzuki} \&
  {Inutsuka}}{2014}]{suzuki14}
{Suzuki} T.~K.,  {Inutsuka} S.-i.,  2014, \mn@doi [\apj]
  {10.1088/0004-637X/784/2/121}, \href
  {https://ui.adsabs.harvard.edu/abs/2014ApJ...784..121S} {784, 121}

\bibitem[\protect\citeauthoryear{{Suzuki}, {Ogihara}, {Morbidelli}, {Crida}  \&
  {Guillot}}{{Suzuki} et~al.}{2016}]{suzuki_etal16}
{Suzuki} T.~K.,  {Ogihara} M.,  {Morbidelli} A.,  {Crida} A.,   {Guillot} T.,
  2016, \mn@doi [\aap] {10.1051/0004-6361/201628955}, \href
  {https://ui.adsabs.harvard.edu/abs/2016A&A...596A..74S} {596, A74}

\bibitem[\protect\citeauthoryear{{Takahashi} \& {Inutsuka}}{{Takahashi} \&
  {Inutsuka}}{2014}]{ts14}
{Takahashi} S.~Z.,  {Inutsuka} S.-i.,  2014, \mn@doi [\apj]
  {10.1088/0004-637X/794/1/55}, \href
  {https://ui.adsabs.harvard.edu/abs/2014ApJ...794...55T} {794, 55}

\bibitem[\protect\citeauthoryear{{Takasao}, {Tomida}, {Iwasaki}  \&
  {Suzuki}}{{Takasao} et~al.}{2018}]{takasao_etal18}
{Takasao} S.,  {Tomida} K.,  {Iwasaki} K.,   {Suzuki} T.~K.,  2018, \mn@doi
  [\apj] {10.3847/1538-4357/aab5b3}, \href
  {https://ui.adsabs.harvard.edu/abs/2018ApJ...857....4T} {857, 4}

\bibitem[\protect\citeauthoryear{{Teague} \& {Loomis}}{{Teague} \&
  {Loomis}}{2020}]{teague_etal20}
{Teague} R.,  {Loomis} R.,  2020, \mn@doi [\apj] {10.3847/1538-4357/aba956},
  \href {https://ui.adsabs.harvard.edu/abs/2020ApJ...899..157T} {899, 157}

\bibitem[\protect\citeauthoryear{{Teague}, {Bae}, {Bergin}, {Birnstiel}  \&
  {Foreman-Mackey}}{{Teague} et~al.}{2018}]{teague_etal18}
{Teague} R.,  {Bae} J.,  {Bergin} E.~A.,  {Birnstiel} T.,   {Foreman-Mackey}
  D.,  2018, \mn@doi [\apjl] {10.3847/2041-8213/aac6d7}, \href
  {https://ui.adsabs.harvard.edu/abs/2018ApJ...860L..12T} {860, L12}

\bibitem[\protect\citeauthoryear{{Tominaga}, {Inutsuka}  \&
  {Takahashi}}{{Tominaga} et~al.}{2018}]{tominaga_etal18}
{Tominaga} R.~T.,  {Inutsuka} S.-i.,   {Takahashi} S.~Z.,  2018, \mn@doi
  [\pasj] {10.1093/pasj/psx143}, \href
  {https://ui.adsabs.harvard.edu/abs/2018PASJ...70....3T} {70, 3}

\bibitem[\protect\citeauthoryear{{Tominaga}, {Takahashi}  \&
  {Inutsuka}}{{Tominaga} et~al.}{2019}]{tominaga_etal19}
{Tominaga} R.~T.,  {Takahashi} S.~Z.,   {Inutsuka} S.-i.,  2019, \mn@doi [\apj]
  {10.3847/1538-4357/ab25ea}, \href
  {https://ui.adsabs.harvard.edu/abs/2019ApJ...881...53T} {881, 53}

\bibitem[\protect\citeauthoryear{{Turner}, {Fromang}, {Gammie}, {Klahr},
  {Lesur}, {Wardle}  \& {Bai}}{{Turner} et~al.}{2014}]{turner_etal14}
{Turner} N.~J.,  {Fromang} S.,  {Gammie} C.,  {Klahr} H.,  {Lesur} G.,
  {Wardle} M.,   {Bai} X.~N.,  2014, in {Beuther} H.,  {Klessen} R.~S.,
  {Dullemond} C.~P.,   {Henning} T.,  eds, Protostars and Planets VI. p.~411
  (\mn@eprint {arXiv} {1401.7306}),
  \mn@doi{10.2458/azu_uapress_9780816531240-ch018}

\bibitem[\protect\citeauthoryear{{Varni{\`e}re} \& {Tagger}}{{Varni{\`e}re} \&
  {Tagger}}{2006}]{vt06}
{Varni{\`e}re} P.,  {Tagger} M.,  2006, \mn@doi [\aap]
  {10.1051/0004-6361:200500226}, \href
  {https://ui.adsabs.harvard.edu/abs/2006A&A...446L..13V} {446, L13}

\bibitem[\protect\citeauthoryear{{Vlemmings} et~al.,}{{Vlemmings}
  et~al.}{2019}]{vlemmings_etal19}
{Vlemmings} W.~H.~T.,  et~al., 2019, \mn@doi [\aap]
  {10.1051/0004-6361/201935459}, \href
  {https://ui.adsabs.harvard.edu/abs/2019A&A...624L...7V} {624, L7}

\bibitem[\protect\citeauthoryear{{Walsh}, {Millar}  \& {Nomura}}{{Walsh}
  et~al.}{2010}]{walsh_etal10}
{Walsh} C.,  {Millar} T.~J.,   {Nomura} H.,  2010, \mn@doi [\apj]
  {10.1088/0004-637X/722/2/1607}, \href
  {https://ui.adsabs.harvard.edu/abs/2010ApJ...722.1607W} {722, 1607}

\bibitem[\protect\citeauthoryear{{Wang}, {Bai}  \& {Goodman}}{{Wang}
  et~al.}{2019}]{wang_etal19}
{Wang} L.,  {Bai} X.-N.,   {Goodman} J.,  2019, \mn@doi [\apj]
  {10.3847/1538-4357/ab06fd}, \href
  {https://ui.adsabs.harvard.edu/abs/2019ApJ...874...90W} {874, 90}

\bibitem[\protect\citeauthoryear{{Wardle}}{{Wardle}}{2007}]{wardle07}
{Wardle} M.,  2007, \mn@doi [\apss] {10.1007/s10509-007-9575-8}, \href
  {https://ui.adsabs.harvard.edu/abs/2007Ap&SS.311...35W} {311, 35}

\bibitem[\protect\citeauthoryear{{Weidenschilling}}{{Weidenschilling}}{1977}]{weidenschilling77}
{Weidenschilling} S.~J.,  1977, \mn@doi [\apss] {10.1007/BF00642464}, \href
  {https://ui.adsabs.harvard.edu/abs/1977Ap&SS..51..153W} {51, 153}

\bibitem[\protect\citeauthoryear{Weiss, Bai  \& Fu}{Weiss et~al.}{2021}]{wbf21}
Weiss B.~P.,  Bai X.-N.,   Fu R.~R.,  2021, \mn@doi [Science Advances]
  {10.1126/sciadv.aba5967}, 7

\bibitem[\protect\citeauthoryear{{W{\"u}nsch}, {Klahr}  \&
  {R{\'o}{\.z}yczka}}{{W{\"u}nsch} et~al.}{2005}]{wunsch05}
{W{\"u}nsch} R.,  {Klahr} H.,   {R{\'o}{\.z}yczka} M.,  2005, \mn@doi [\mnras]
  {10.1111/j.1365-2966.2005.09319.x}, \href
  {https://ui.adsabs.harvard.edu/abs/2005MNRAS.362..361W} {362, 361}

\bibitem[\protect\citeauthoryear{{Youdin}}{{Youdin}}{2011}]{youdin11}
{Youdin} A.~N.,  2011, \mn@doi [\apj] {10.1088/0004-637X/731/2/99}, \href
  {https://ui.adsabs.harvard.edu/abs/2011ApJ...731...99Y} {731, 99}

\bibitem[\protect\citeauthoryear{{Zanni}, {Ferrari}, {Rosner}, {Bodo}  \&
  {Massaglia}}{{Zanni} et~al.}{2007}]{zanni_etal07}
{Zanni} C.,  {Ferrari} A.,  {Rosner} R.,  {Bodo} G.,   {Massaglia} S.,  2007,
  \mn@doi [\aap] {10.1051/0004-6361:20066400}, \href
  {https://ui.adsabs.harvard.edu/abs/2007A&A...469..811Z} {469, 811}

\bibitem[\protect\citeauthoryear{{Zhang}, {Blake}  \& {Bergin}}{{Zhang}
  et~al.}{2015}]{zhang_etal15}
{Zhang} K.,  {Blake} G.~A.,   {Bergin} E.~A.,  2015, \mn@doi [\apjl]
  {10.1088/2041-8205/806/1/L7}, \href
  {https://ui.adsabs.harvard.edu/abs/2015ApJ...806L...7Z} {806, L7}

\bibitem[\protect\citeauthoryear{{Zhao} et~al.,}{{Zhao} et~al.}{2020}]{zhao+20}
{Zhao} B.,  et~al., 2020, \mn@doi [\ssr] {10.1007/s11214-020-00664-z}, \href
  {https://ui.adsabs.harvard.edu/abs/2020SSRv..216...43Z} {216, 43}

\bibitem[\protect\citeauthoryear{{Zhu} \& {Stone}}{{Zhu} \&
  {Stone}}{2018}]{zhu18}
{Zhu} Z.,  {Stone} J.~M.,  2018, \mn@doi [\apj] {10.3847/1538-4357/aaafc9},
  \href {https://ui.adsabs.harvard.edu/abs/2018ApJ...857...34Z} {857, 34}

\bibitem[\protect\citeauthoryear{{Zhu}, {Nelson}, {Hartmann}, {Espaillat}  \&
  {Calvet}}{{Zhu} et~al.}{2011}]{zhu_etal11}
{Zhu} Z.,  {Nelson} R.~P.,  {Hartmann} L.,  {Espaillat} C.,   {Calvet} N.,
  2011, \mn@doi [\apj] {10.1088/0004-637X/729/1/47}, \href
  {https://ui.adsabs.harvard.edu/abs/2011ApJ...729...47Z} {729, 47}

\bibitem[\protect\citeauthoryear{{van Boekel} et~al.,}{{van Boekel}
  et~al.}{2017}]{vanboekel_etal17}
{van Boekel} R.,  et~al., 2017, \mn@doi [\apj] {10.3847/1538-4357/aa5d68},
  \href {https://ui.adsabs.harvard.edu/abs/2017ApJ...837..132V} {837, 132}

\bibitem[\protect\citeauthoryear{{van der Marel} et~al.,}{{van der Marel}
  et~al.}{2013}]{vdmarel_etal13}
{van der Marel} N.,  et~al., 2013, \mn@doi [Science] {10.1126/science.1236770},
  \href {https://ui.adsabs.harvard.edu/abs/2013Sci...340.1199V} {340, 1199}

\bibitem[\protect\citeauthoryear{{van der Marel} et~al.,}{{van der Marel}
  et~al.}{2020}]{vdmarel_etal20}
{van der Marel} N.,  et~al., 2020, arXiv e-prints, \href
  {https://ui.adsabs.harvard.edu/abs/2020arXiv201010568V} {p. arXiv:2010.10568}

\makeatother
\end{thebibliography}

\end{document}